\documentclass[12pt,a4paper]{article} 

\usepackage{amsmath,a4,amssymb,epsfig} 
\makeatletter

\def\section{\@startsection{section}{1}{\z@}{-3.25ex plus -1ex minus
    -.2ex}{1.5ex plus .2ex}{\normalfont\bfseries}}
\def\subsection{\@startsection{subsection}{1}{\z@}{-3.25ex plus -1ex
    minus -.2ex}{1.5ex plus .2ex}{\normalfont\itshape}}

\@addtoreset{equation}{section}

\makeatother

\allowdisplaybreaks[4]

\newtheorem{lem}{Lemma} 

\newtheorem{thm}[lem]{Theorem}

\renewcommand{\title}[1]{\vspace{10mm}\noindent{\Large{\bf #1}}\vspace{8mm}}
\newcommand{\authors}[1]{\noindent{\large #1}\vspace{3mm}}
\newcommand{\address}[1]{{\itshape #1\vspace{2mm}}}

\begin{document}

\begin{titlepage}

\hspace*{\fill}hep-th/0307017 

\begin{center}
  
  \title{Renormalisation of $\phi^4$-theory on noncommutative
    $\mathbb{R}^2$ \\[1mm] in the matrix base}

\authors{Harald {\sc Grosse}$^1$ and Raimar {\sc Wulkenhaar}$^2$}

\address{$^{1}$\,Institut f\"ur Theoretische Physik, Universit\"at Wien\\
Boltzmanngasse 5, A-1090 Wien, Austria}

\address{$^{2}$\,Max-Planck-Institut f\"ur Mathematik in den
  Naturwissenschaften\\ 
Inselstra\ss{}e 22-26, D-04103 Leipzig, Germany}

\footnotetext[1]{harald.grosse@univie.ac.at}
\footnotetext[2]{Schloe\ss{}mann fellow, raimar.wulkenhaar@mis.mpg.de}

\vskip 2cm

\textbf{Abstract}
\vskip 3mm

\begin{minipage}{14cm}
  As a first application of our renormalisation group approach to
  non-local matrix models [hep-th/0305066], we prove (super-)
  renormalisability of Euclidean two-dimensional noncommutative
  $\phi^4$-theory. It is widely believed that this model is
  renormalisable in momentum space arguing that there would be
  logarithmic UV/IR-divergences only. Although momentum space Feynman
  graphs can indeed be computed to any loop order, the logarithmic
  UV/IR-divergence appears in the renormalised two-point function---a
  hint that the renormalisation is not completed. In particular, it is
  impossible to define the squared mass as the value of the two-point
  function at vanishing momentum. In contrast, in our matrix approach
  the renormalised $N$-point functions are bounded everywhere and
  nevertheless rely on adjusting the mass only. We achieve this by
  introducing into the cut-off model a translation-invariance breaking
  regulator which is scaled to zero with the removal of the cut-off.
  The na\"{\i}ve treatment without regulator would not lead to a
  renormalised theory.
\end{minipage}

\end{center}

\end{titlepage}

\section{Introduction}

In spite of enormous efforts, the renormalisation of quantum field
theories on the noncommutative $\mathbb{R}^D$ is not achieved. These
models show a phenomenon called \emph{UV/IR-mixing}
\cite{Minwalla:1999px} which was analysed to all orders by Chepelev
and Roiban \cite{Chepelev:1999tt,Chepelev:2000hm}. The conclusion of
the power-counting theorem is that, in general, field theories on
noncommutative $\mathbb{R}^D$ are not renormalisable if their
commutative counterparts are worse than logarithmically divergent. The
situation is better for models with at most logarithmic divergences.
Applying the power-counting analysis to the real $\phi^4$-model on
noncommutative $\mathbb{R}^2$, one finds ``that the divergences from
all connected Green's functions at non-exceptional external momenta
can be removed in the counter-term approach'' (literally quoted from
\cite[\S 4.3]{Chepelev:2000hm}). The problem is, however, that
non-exceptional momenta can become arbitrarily close to exceptional
momenta so that the renormalised Green's functions are
\emph{unbounded}. Although one can probably live with that, it is not
a desired feature of a quantum field theory.

We have elaborated in \cite{Grosse:2003aj} the Wilson-Polchinski
renormalisation group approach \cite{Wilson:1973jj, Polchinski:1983gv}
for dynamical matrix models where the propagator is neither diagonal
nor constant. We have derived a power-counting theorem for ribbon
graphs by solving the exact renormalisation group equation
perturbatively. The power-counting degree of divergence of a ribbon
graph is determined by its topology and the asymptotic behaviour of
the cut-off propagator.  Our motivation was to provide a
renormalisation scheme for very general noncommutative field theories,
because the typical noncommutative geometries are matrix geometries.
The noncommutative $\mathbb{R}^D$ is no exception as there exists a
matrix base \cite{Gracia-Bondia:1987kw} in which the $\star$-product
interaction becomes the trace of an ordinary product of matrices. The
propagator becomes complicated in the matrix base but as we show in
this paper, the difficulties can be overcome.

In \cite{Grosse:2003aj} we have only completed the first (but most essential)
step of Polchinski's approach \cite{Polchinski:1983gv}, namely the integration
of the flow equation between a finite initial scale $\Lambda_0$ and the
renormalisation scale $\Lambda_R$. In order to prove renormalisability the
limit $\Lambda_0\to\infty$ has to be taken. This step is model dependent. We
focus in this paper on the real $\phi^4$-theory on noncommutative
$\mathbb{R}^2$. The na\"{\i}ve idea would be to take the standard
$\phi^4$-action at the initial scale $\Lambda_0$, with $\Lambda_0$-dependent
bare mass to be adjusted such that at $\Lambda_R$ it is scaled down to the
renormalised mass.  Unfortunately, this does not work. In the limit
$\Lambda_0\to\infty$ one obtains an unbounded power-counting degree of
divergence for the ribbon graphs. The solution is the observation that the
cut-off action at $\Lambda_0$ is (due to the cut-off) not translation
invariant. We are therefore free to break the translational symmetry of the
action at $\Lambda_0$ even more by adding a harmonic oscillator potential for
the fields $\phi$. We prove that there exists a $\Lambda_0$-dependence of the
oscillator frequency $\Omega$ with $\lim_{\Lambda_0 \to \infty} \Omega=0$ such
that the effective action at $\Lambda_R$ is convergent (and thus bounded)
order by order in the coupling constant in the limit $\Lambda_0 \to \infty$.
This means that the partition function of the original (translation-invariant)
$\phi^4$-model without cut-off and with suitable divergent bare mass is solved
by Feynman graphs with propagators cut-off at $\Lambda_R$ and vertices given
by the bounded expansion coefficients of the effective action at $\Lambda_R$.
Hence, this model is renormalisable, and there is no problem with exceptional
configurations.

We are optimistic that in the same way we can renormalise the
$\phi^4$-model on noncommutative $\mathbb{R}^4$ \cite{gw3}.

\section{$\phi^4$-theory on noncommutative $\mathbb{R}^D$}

\subsection{The regularised action in the matrix base}

The noncommutative $\mathbb{R}^D$, $D=2,4,6,\dots$, is defined as the
algebra $\mathbb{R}^D_\theta$ which as a vector space is given by
the space $\mathcal{S}(\mathbb{R}^D)$ of (complex-valued) Schwartz
class functions of rapid decay, equipped with the multiplication rule
\cite{Gracia-Bondia:1987kw}
\begin{align}
  (a\star b)(x) &= \int \frac{d^Dk}{(2\pi)^D} \int d^D y \;
  a(x{+}\tfrac{1}{2} \theta {\cdot} k)\, b(x{+}y)\,
  \mathrm{e}^{\mathrm{i} k \cdot y}\;,
\label{starprod}
\\*
& (\theta {\cdot} k)^\mu = \theta^{\mu\nu} k_\nu\;,\quad k{\cdot}y =
k_\mu y^\mu\;,\quad \theta^{\mu\nu}=-\theta^{\nu\mu}\;.  \nonumber
\end{align}
The entries $\theta^{\mu\nu}$ in (\ref{starprod}) have the dimension
of an area. 

We are going to study a regularised $\phi^4$-theory on
$\mathbb{R}^D_\theta$ defined by the action
\begin{align}
  S_D[\phi] &= \int d^Dx \Big( \frac{1}{2} g^{\mu\nu}
  \big(\partial_\mu \phi \star \partial_\nu \phi +4 \Omega^2
  ((\theta^{-1})_{\mu\rho} x^\rho \phi ) \star
  ((\theta^{-1})_{\nu\sigma} x^\sigma \phi) \big) + \frac{1}{2}
  \mu_0^2 \,\phi \star \phi \nonumber
  \\
  &\hspace*{10em} + \frac{\lambda}{4!} \phi \star \phi \star \phi \star
  \phi\Big) \;,
\label{action}
\end{align}
which is given by adding a harmonic oscillator potential to the
standard $\phi^4$-action. The potential beaks translation invariance.
We shall learn that the renormalisation of standard $\phi^4$-theory
has to be performed along a path of actions (\ref{action}). 

Our goal is to write the classical action (\ref{action}) in an adapted
base.  We place ourselves into a coordinate system in which $\theta$
has in $D$ dimensions the form
\begin{align}
  \theta_{\mu\nu} &=\left(\begin{array}{cccc}
      \boldsymbol{\theta}_1 & 0 & \dots & 0 \\
      0 & \boldsymbol{\theta}_2 & \dots & 0 \\
      \vdots & \vdots & \ddots & \vdots \\
      0 & 0 & \dots & \boldsymbol{\theta}_{\frac{D}{2}}
\end{array}\right)\;, &
\boldsymbol{\theta}_i = \left(\begin{array}{cc}
    0 & \theta_i \\
    - \theta_i & 0
\end{array}\right)\;.
\end{align}
Now an adapted base of $\mathbb{R}^D_\theta$ is
\begin{align}
  b_{mn}(x) &= f_{m_1n_1}(x_1,x_2) \,f_{m_2n_2}(x_3,x_4) \dots
  f_{m_{D/2}n_{D/2}}(x_{D-1},x_D) \;,
\label{bbas}
\\*
& m=(m_1,m_2,\dots,m_{D/2})\in \mathbb{N}^{\frac{D}{2}}\,,~
n=(n_1,n_2,\dots,n_{D/2})\in \mathbb{N}^{\frac{D}{2}}\;, \nonumber
\end{align}
where the base $f_{mn}(x_1,x_2) \in \mathbb{R}^2_\theta$ is
introduced in (\ref{fmn}) in Appendix~\ref{appA}.

The advantage of this base is that the $\star$-product
(\ref{starprod}) is represented by a product (\ref{fprodmat}) of
infinite matrices and that the multiplication by $x^\rho$ is easy
to realise. This means that expanding the fields according
to
\begin{align}
  \phi(x)=\sum_{m,n \in \mathbb{N}^{\frac{D}{2}}} \phi_{mn} b_{mn}(x) \;,
\label{phimat}
\end{align}
the interaction term $\phi\star\phi\star\phi\star\phi$ in
(\ref{action}) becomes very simple. The price for this simplification
is, however, that the kinetic term given by the first line in
(\ref{action}) becomes very complicated. In \cite{Grosse:2003aj} we
have extended the first stept of Polchinski's renormalisation proof
\cite{Polchinski:1983gv} of commutative $\phi^4$-theory to a
renormalisation method suited for dynamical matrix models with
arbitrary non-diagonal and non-constant propagators. The kinetic term
of the action (\ref{action}) fits precisely into the scope of
\cite{Grosse:2003aj}.

\subsection{Computation of the propagator in the two-dimensional case}

For the remainder of this paper we restrict ourselves to $D=2$
dimensions. The four-dimensional case will be treated elsewhere
\cite{gw3}.  Using the formulae collected in Appendix~\ref{appA} we
first calculate the kinetic term in two dimensions:
\begin{align}
G_{mn;kl} 
&:= \int \frac{d^2x}{2\pi\theta_1} \, \Big(\big(\partial_1 f_{mn}
  \star \partial_1 f_{kl} + \partial_2 f_{mn} \star \partial_2 f_{kl}
\nonumber
\\*
&\qquad \qquad + \frac{4 \Omega^2}{\theta^2} \big( (x_1 f_{mn}) \star
  (x_1 f_{kl}) + (x_2 f_{mn}) \star (x_2 f_{kl}) \big) + \mu_0^2\,
  f_{mn}\star f_{kl} \Big) \nonumber
\\
& =\int \frac{d^2x}{2\pi\theta_1} \,
\Big( \frac{1{+}\Omega^2}{\theta_1^2} f_{mn}
  \star (a\star \bar{a}+\bar{a} \star a) \star f_{kl} 
+\frac{1{+}\Omega^2}{\theta_1^2}   f_{kl} \star (a\star \bar{a}
+\bar{a} \star a) \star f_{mn} \nonumber
\\*
  &\qquad -\frac{2(1{+}\Omega^2)}{\theta_1^2} 
f_{mn} \star a \star f_{kl} \star  \bar{a} 
-\frac{2(1{+}\Omega^2)}{\theta_1^2} f_{kl} \star a \star f_{mn} \star \bar{a}
  + \mu_0^2 \, f_{mn}\star f_{kl} \Big) \nonumber
\\*
&=\Big(\mu_0^2+ \frac{2(1{+}\Omega^2)}{\theta_1}(m{+}n{+}1)
%+ \frac{4 \Omega}{\theta_1}(n{-}m)
\Big) \delta_{nk} \delta_{ml} \nonumber
\\*
&\qquad - \frac{2(1{-}\Omega^2)}{\theta_1} \sqrt{(n{+}1)(m{+}1)}
  \delta_{n+1,k}\delta_{m+1,l} - \frac{2(1{-}\Omega^2)}{\theta_1} \sqrt{nm}
  \delta_{n-1,k}\delta_{m-1,l}\;.
\label{G2Dcal}
\end{align}
Defining
\begin{align}
  \mu^2 &= \frac{2(1+\Omega^2)}{\theta_1} \;, & \sqrt{\omega} &=
  \frac{1-\Omega^2}{1+\Omega^2} \;,
\label{mu}
\end{align}
with $-1 < \sqrt{\omega} \leq 1$, we can rewrite (\ref{G2Dcal}) as
\begin{align}
  G_{mn;kl} &= \big(\mu_0^2{+} (n{+}m{+}1)\mu^2
%+ (n{-}m) \sqrt{1{-}\omega}\,\mu^2 
  \big) \delta_{nk} \delta_{ml} \nonumber
  \\*
  & - \mu^2 \sqrt{\omega\, (n{+}1)(m{+}1)}\,
  \delta_{n+1,k}\delta_{m+1,l} - \mu^2 \sqrt{\omega\,nm}\,
  \delta_{n-1,k} \delta_{m-1,l}\,.
\label{G2D}
\end{align}
Now the action (\ref{action}) takes the form
\begin{align}
  S_2[\phi] &= 2\pi\theta_1 \sum_{m,n,k,l} \Big( \frac{1}{2} 
\phi_{mn} G_{mn;kl} \phi_{kl} + \frac{\lambda}{4!} 
\phi_{mn} \phi_{nk} \phi_{kl} \phi_{lm}\Big)\,.
\end{align}

Next we are going to invert $G_{mn;kl}$, i.e.\ we solve in the
two-dimensional case
\begin{align}
  \sum_{k,l=0}^\infty G_{mn;kl} \Delta_{lk;sr} = \sum_{k,l=0}^\infty
  \Delta_{nm;lk} G_{kl;rs} = \delta_{mr} \delta_{ns}\,.
\label{GD}
\end{align}
The indices $m,n,k,l$ of each term contributing to (\ref{G2D}) are
restricted by
\begin{align}
  m+k=n+l\;.
\label{mnkl}
\end{align}
Since the same relation is induced for the propagator $\Delta_{lk;nm}$ as
well, the problem to solve (\ref{GD}) factorises into the independent
equations
\begin{align}
  \sum_{l=0}^\infty G_{m,m+\alpha;l+\alpha,l}
  \Delta_{l,l+\alpha;r+\alpha,r} = \sum_{l=0}^\infty
  \Delta_{m+\alpha,m;l,l+\alpha} G_{l+\alpha,l;r,r+\alpha} =
  \delta_{mr} \,.
\label{GDv}
\end{align}
We define $\Delta_{mn;kl}=0$ and $G_{mn;kl}=0$ if one of the indices
$m,n,k,l$ is negative. For each $\alpha$ we have to invert an
infinite square matrix.  We therefore introduce a cut-off
$\mathcal{N}$ with $0\leq m,n,k,l,r,s < \mathcal{N}$ above. Our
strategy is to diagonalise the massless kinetic term
\begin{align}
  G^{(\mathcal{N})}_{m,m+\alpha;l+\alpha,l}\Big|_{\mu_0=0} & =\mu^2
  \sum_{i=1}^{\mathcal{N}} U_{m+1,i}^{(\mathcal{N},\alpha,\omega)}\,
  v_i \, U_{i,l+1}^{(\mathcal{N},\alpha,\omega)*} \;,
\nonumber
\\*
\delta_{ml} &= \sum_{i=1}^{\mathcal{N}} U_{mi}^{(\mathcal{N},\alpha,\omega)}
  U_{il}^{(\mathcal{N},\alpha,\omega)*} =\sum_i
  U_{mi}^{(\mathcal{N},\alpha,\omega)*} U_{il}^{(\mathcal{N},\alpha,\omega)} 
\;.
\label{GUlU}
\end{align}
To see what result we can expect let us consider the eigenvalue
problem of $\mathcal{N}=4+\alpha$ and $\alpha \geq 0$:
\begin{align}
  &G^{(4)}_{m,m+\alpha;l+\alpha,l}\Big|_{\mu_0=0} -v \mu^2 \delta_{ml}^{(4)}
  \nonumber
  \\*
  &= \mu^2 \left(\begin{array}{ccccc}
      \alpha{+}1{-}v & -\sqrt{1(\alpha{+}1)\omega} & 0 & 0 \\
      -\sqrt{1(\alpha{+}1)\omega} & \alpha{+}3{-}v &
      -\sqrt{2(\alpha{+}2)\omega} & 0  \\
      0 & -\sqrt{2(\alpha{+}2)\omega} & \alpha{+}5{-}v &
      -\sqrt{3(\alpha{+}3)\omega} \\
      0 & 0 & -\sqrt{3(\alpha{+}3)\omega} & \alpha{+}7{-}v
\end{array}
\right)_{\!\!m+1,l+l} \nonumber
\\
&= \mu^2 \left(\begin{array}{ccccc}
    \sqrt{\alpha{+}1}\sqrt{A^{\alpha,\omega}_1(v)} & 0 & 0 & 0  \\
    -\sqrt{\frac{1\, \omega }{A^{\alpha,\omega}_1(v)}} &
    \sqrt{\alpha{+}2}\sqrt{A^{\alpha,\omega}_2(v)} & 0 & 0
    \\
    0 & -\sqrt{\frac{2 \omega}{A^{\alpha,\omega}_2(v)}} &
    \sqrt{\alpha{+}3}\sqrt{A^{\alpha,\omega}_3(v)} & 0
    \\
    0 & 0 & -\sqrt{\frac{3\omega}{A^{\alpha,\omega}_3(v)}} &
    \sqrt{\alpha{+}4}\sqrt{A^{\alpha,\omega}_4(v)}
\end{array}\right)
\nonumber
\\*
& \qquad \times \left(\begin{array}{ccccc}
    \sqrt{\alpha{+}1}\sqrt{A^{\alpha,\omega}_1(v)}
    &-\sqrt{\frac{1\,\omega}{A^{\alpha,\omega}_1(v)}} & 0 & 0
    \\
    0 & \sqrt{\alpha{+}2}\sqrt{A^{\alpha,\omega}_2(v)} &
    -\sqrt{\frac{2\omega}{A^{\alpha,\omega}_2(v)}} & 0
    \\
    0 & 0 & \sqrt{\alpha{+}3}\sqrt{A^{\alpha,\omega}_3(v)} &
    -\sqrt{\frac{3\omega}{A^{\alpha,\omega}_3(v)}}
    \\
    0 & 0 & 0 & \sqrt{\alpha{+}4}\sqrt{A^{\alpha,\omega}_4(v)}
\end{array}\right),
\label{G2N}
\end{align}
where
\begin{align}
  A^{\alpha,\omega}_n(v) &:= \frac{1}{\alpha{+}n}\Big(
  \alpha+2n-1-v
  -\frac{(n{-}1)\omega}{A^{\alpha,\omega}_{n-1}(v)}\Big)\;, \qquad
  n\geq 1\;.
\label{Anla}
\end{align}
Note that $0\leq \omega:=(\sqrt{\omega})^2 \leq 1$. With the ansatz
\begin{align}
  A^{\alpha,\omega}_n(v) &= \frac{n}{\alpha{+}n} \frac{
    L_n^{\alpha,\omega}(v)}{L_{n-1}^{\alpha,\omega}(v)}\;,\qquad
  L^{\alpha,\omega}_0(v) \equiv 1\;,
\label{Laguerre}
\end{align}
(\ref{Anla}) can be rewritten as
\begin{align}
  0 = n L^{\alpha,\omega}_n (v) - (\alpha+2n-1-v)
  L^{\alpha,\omega}_{n-1}(v) + \omega(\alpha+n-1)
  L^{\alpha,\omega}_{n-2}(v) \;.
\label{Laguerrerec}
\end{align}
For $\omega=1$ we recognise this relation as the recursion relation of
Laguerre polynomials \cite[\S 8.971.6]{GR}. We thus denote the $
L^{\alpha,\omega}_n(v)$ as \emph{deformed Laguerre polynomials},
with $ L^{\alpha,1}_n(v)\equiv L^\alpha_n(v)$ being the
usual Laguerre polynomials.

At given matrix cut-off $\mathcal{N}$ it follows from (\ref{G2N}) and
(\ref{Laguerre}) that the eigenvalues $v_i$ are the zeroes of
the deformed Laguerre polynomial $L^{\alpha,\omega}_{\mathcal{N}}$:
\begin{align}
  L^{\alpha,\omega}_{\mathcal{N}}\big(v_i^{(\mathcal{N},\alpha,\omega)}\big)
  &=0\;, \qquad i=1,\dots,\mathcal{N}\;,
  \\
  U^{(\mathcal{N},\alpha,\omega)}_{ji} = U^{(\mathcal{N},\alpha,\omega)*}_{ij}
  &=\sqrt{\frac{\Gamma(\alpha{+}\mathcal{N}) \Gamma(j)
      \omega^{\mathcal{N}}}{ \Gamma(\alpha{+}j) \Gamma(\mathcal{N})
      \omega^j} }
  \frac{L_{j-1}^{\alpha,\omega}(v_i^{(\mathcal{N},\alpha,\omega)})}{
    L_{\mathcal{N}-1}^{\alpha,\omega}(v_i^{(\mathcal{N},\alpha,\omega)})}
  U_{\mathcal{N}i}^{(\mathcal{N},\alpha,\omega)} \nonumber
  \\*
  &= \frac{\sqrt{\frac{\Gamma(j)}{\omega^{j-1}\Gamma(\alpha+j)}}
    L_{j-1}^{\alpha,\omega} (v_i^{(\mathcal{N},\alpha,\omega)})}{
    \sqrt{\sum_{h=1}^\mathcal{N}
      \frac{\Gamma(h)}{\omega^{h-1}\Gamma(\alpha+h)}
      \big(L_{h-1}^{\alpha,\omega}(v_i^{(\mathcal{N},\alpha,\omega)})\big)^2
    }} \;, \quad j=1,\dots,\mathcal{N}\;.
\label{UNa}
\end{align}

Inserting (\ref{UNa}) into (\ref{GUlU}) and (\ref{GDv}) we obtain for
$\alpha=n{-}m=k{-}l\geq 0$ the solutions
\begin{align}
  \delta_{ml}^{(\mathcal{N},\alpha,\omega)} &= \sum_{i=1}^{\mathcal{N}}
  \frac{\sqrt{\frac{m!l!}{\omega^{m+l}\,(m+\alpha)!(l+\alpha)!}}
    L^{\alpha,\omega}_m(v_i^{(\mathcal{N},\alpha,\omega)})
    L^{\alpha,\omega}_l(v_i^{(\mathcal{N},\alpha,\omega)}) }{
    \sum_{h=0}^{\mathcal{N}-1} \frac{h!}{ \omega^h(\alpha+h)!}
    \big(L_{h}^{\alpha,\omega}(v_i^{(\mathcal{N},\alpha,\omega)})\big)^2
  } \;,
\label{orthN} 
\\
G_{m,m+\alpha;l+\alpha,l}^{(\mathcal{N},\alpha,\omega)} 
&= \sum_{i=1}^{\mathcal{N}}
  \frac{\sqrt{\frac{m!l!}{\omega^{m+l}\,(m+\alpha)!(l+\alpha)!}}
    L^{\alpha,\omega}_m(v_i^{(\mathcal{N},\alpha,\omega)})
    L^{\alpha,\omega}_l(v_i^{(\mathcal{N},\alpha,\omega)}) }{
    \sum_{h=0}^{\mathcal{N}-1} \frac{h!}{ \omega^h(\alpha+h)!}
    \big(L_{h}^{\alpha,\omega}(v_i^{(\mathcal{N},\alpha,\omega)})\big)^2}
\Big( \mu_0^2 + \mu^2 v_i^{(\mathcal{N},\alpha,\omega)}\Big)\;,
\label{GN}
\\
\Delta_{m+\alpha,m;l,l+\alpha}^{(\mathcal{N},\alpha,\omega)} 
&= \sum_{i=1}^{\mathcal{N}}
  \frac{\sqrt{\frac{m!l!}{\omega^{m+l}\,(m+\alpha)!(l+\alpha)!}}
    L^{\alpha,\omega}_m(v_i^{(\mathcal{N},\alpha,\omega)})
    L^{\alpha,\omega}_l(v_i^{(\mathcal{N},\alpha,\omega)}) }{
    \sum_{h=0}^{\mathcal{N}-1} \frac{h!}{ \omega^h(\alpha+h)!}
    \big(L_{h}^{\alpha,\omega}(v_i^{(\mathcal{N},\alpha,\omega)})\big)^2}
\;\frac{1}{\mu_0^2 + \mu^2 v_i^{(\mathcal{N},\alpha,\omega)}}\;.
\label{DeltaN}
\end{align}
Since the kinetic term (\ref{G2D}) is symmetric in $(m\leftrightarrow
n,\,k\leftrightarrow l)$, we obtain the analogue of (\ref{GN}) and
(\ref{DeltaN}) in the case $\alpha=n{-}m=k{-}l\leq 0$ by exchanging
$(m\leftrightarrow n,\,k\leftrightarrow l)$.  Note that the recursion
relation (\ref{Laguerrerec}) and the orthogonality (\ref{orthN}) yield
directly the kinetic term (\ref{G2D}).

\subsection{Remarks on the limit $\mathcal{N}\to \infty$}

Now we have to take the limit $\mathcal{N}\to \infty$, which can be
done explicitly for $\omega=0$ and $\omega=1$. For $\omega=0$ we can
invert (\ref{G2D}) directly:
\begin{align}
  \Delta^{(\omega=0)}_{nm;lk}=
  \frac{\delta_{ml}\delta_{nk}}{\mu_0^2 + \mu^2 (m+n+1)}\;.
\end{align}
For $\omega=1$ the zeroes $v_i^{(\mathcal{N},\alpha,1)}$ of the
true Laguerre polynomials $L^\alpha_n$ become continuous variables
$v$, and $\big(\sum_{h=0}^\infty
\frac{h!}{(\alpha+h)!}\big(L_h^\alpha(v)\big)^2 \big)^{-1}$ is
promoted to the measure of integration. This measure is identified by
comparison of (\ref{orthN}) with the standard orthogonality relation
\cite[\S 8.904]{GR} of Laguerre polynomials
\begin{align}
  \delta_{ml} &= \int_0^\infty dv\; v^\alpha
  \,\mathrm{e}^{-v} \sqrt{\frac{m!l!}{(m{+}\alpha)!(l{+}\alpha)!}}  
L^{\alpha}_m(v) L^\alpha_l(v) \;.%\qquad n{-}m=k{-}l\geq 0\;.
\label{orth}
\end{align}
We thus have to translate (\ref{DeltaN}) in the limit $\mathcal{N}\to
\infty$ into
\begin{align}
  \Delta^{(\omega=1)}_{nm;lk} &= \int_0^\infty dv\; v^{n-m}
      \,\mathrm{e}^{-v} \sqrt{\frac{m!l!}{n!k!}}  \,
      \frac{L^{n-m}_m(v) \,L^{k-l}_l(v)}{ \mu_0^2+v \mu^2} 
\delta_{m+k,n+l}\;.
\label{Deltaexact}
\end{align}
We have derived the formula (\ref{Deltaexact}) for $n{-}m=k{-}l\geq 0$
only. However, due to the identity
\begin{align}
L^{-\alpha}_{m+\alpha}(v) = \frac{m!}{(m{+}\alpha)!} (-1)^\alpha 
v^\alpha  L^\alpha_m(v) 
\end{align}
it can be transformed into the $(m\leftrightarrow n,l\leftrightarrow
k)$-exchanged form so that (\ref{Deltaexact}) holds actually for any
$n{-}m=k{-}l$.

Introducing a Schwinger parameter and using \cite[\S 7.414.4]{GR} we can
integrate (\ref{Deltaexact}) to
\begin{align}
  \Delta^{(\omega=1)}_{nm;lk} &= \frac{1}{\mu_0^2} \int_0^\infty \!\!dt
  \int_0^\infty dv\; v^{n-m}
  \,\mathrm{e}^{-v(1+\frac{\mu^2}{\mu_0^2}t)-t}
  \sqrt{\frac{m!l!}{n!k!}}  \, L^{n-m}_m(v) \,L^{k-l}_l(v)
  \,\delta_{m+k,n+l} \nonumber
  \\*
  &= \frac{1}{\mu_0^2} \sqrt{\frac{(n{+}l)!}{n!l!}
    \frac{(m{+}k)!}{m!k!}}  
  \,\delta_{m+k,n+l} 
\int_0^\infty \!\!\! dt
  \;\frac{\big(\frac{\mu^2}{\mu_0^2} t\big)^{m+l} \,\mathrm{e}^{-t}
  }{ \big(1{+}\frac{\mu^2}{\mu_0^2}t\big)^{n+l+1}}\,
  F\Big({-}m,{-}l;{-}n{-}l;1{-}\frac{\mu_0^4}{\mu^4 t^2}\Big) \,.
\label{2F1}
\end{align}
Again, due to the property \cite[\S 9.131.1]{GR} of the hypergeometric
function the result (\ref{2F1}) is invariant under the
exchange $m\leftrightarrow n$ and $k\leftrightarrow l$. 

\begin{figure}[ht]
\begin{picture}(120,35)
\put(-20,-150){\epsfig{file=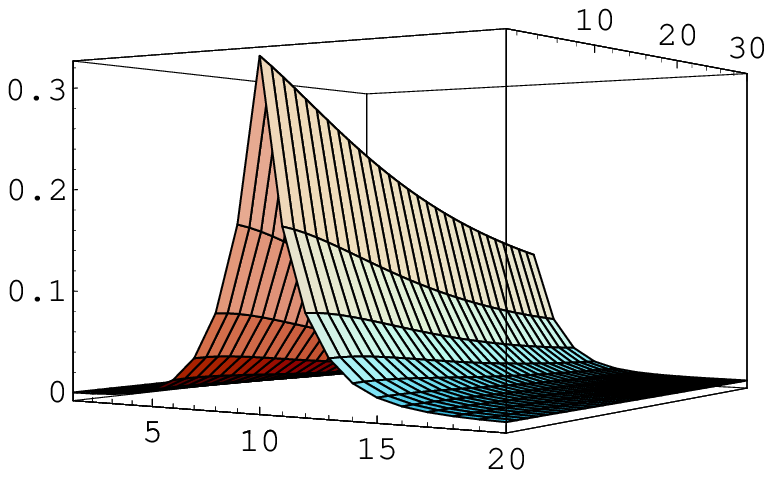,scale=0.7}}
\put(45,-150){\epsfig{file=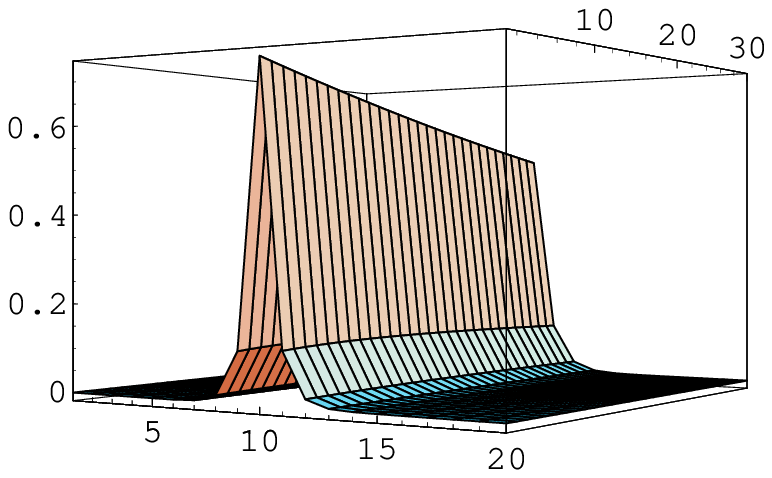,scale=0.7}}
\put(135,10){\vector(0,1){15}}
\put(135,10){\vector(4,-1){12}}
\put(135,10){\vector(3,1){10}}
\put(132,31){\mbox{\scriptsize$\Delta^{(\omega=1)}_{10,10+i;j+i,j}$}}
\put(132,27){\mbox{\scriptsize in units of $\mu_0^{-2}$}}
\put(146,14){\mbox{\scriptsize$i$}}
\put(148,5){\mbox{\scriptsize$j$}}
\put(25,0){\mbox{\scriptsize$\theta_1=10 \mu_0^{-2}$}}
\put(85,0){\mbox{\scriptsize$\theta_1=100 \mu_0^{-2}$}}
\end{picture}
\caption{\label{fig0} The plot of the propagators
  $\Delta_{10,10+i;j+i,j}$ over $i$ and $j$, for two values of $\theta_1$.}
\end{figure}

We recall that in the momentum space version of the $\phi^4$-model,
the interactions contain oscillating phase factors which to our
opinion \cite{Grosse:2003aj} make a Wilson-Polchinski treatment
impossible. Here we use an adapted base which eliminates the phase
factors from the interaction. At first sight it seems that these
oscillations reappear in the propagator via the Laguerre polynomials.
We see, however, from (\ref{2F1}) that this is not the case. The
interpolation of the matrix propagator consists of two monotonous and
apparently smooth parts which are glued together at $\alpha=0$. We
show in Figure~\ref{fig0}% 
how $\Delta_{10,10+i;j+i,j}$ depends on the parameters $i,j$ for the
indices. The monotonous behaviour is perfect for the renormalisation
group approach. One observes that the maximum of $\Delta_{nm;lk}$ for
given (large enough) $n$ is found at $m=n=k=l$. The decay rate of
$\Delta_{nm;lk}$ for increasing indices decides according to
\cite{Grosse:2003aj} about renormalisability. It turns out that
$\Delta_{nm;lk}^{(\omega=1)}$ decays too slowly so that we have to
pass to $\omega<1$. For $\theta_1\to \infty$ one obtains an ordinary
matrix model,
\begin{align}
\lim_{\theta_1 \to \infty} 
\Delta_{nm;lk}=\frac{1}{\mu_0^2} \delta_{ml} \delta_{nk}\;.
\end{align}
This should be compared with \cite{Becchi:2003dg}.

It would be desirable to have an explicit formula as (\ref{2F1}) for
the $\mathcal{N}\to\infty$ limit in case of $\omega< 1$, too. For
that purpose a deeper understanding of the deformed Laguerre
polynomials is indispensable.

\section{The general strategy of renormalisation}

\subsection{Projection to the irrelevant part}

Guided by Wilson's understanding of renormalisation
\cite{Wilson:1973jj} in terms of the scaling of effective Lagrangians,
Polchinski has given a very efficient renormalisation proof of
commutative $\phi^4$-theory in four dimensions
\cite{Polchinski:1983gv}. We have adapted in \cite{Grosse:2003aj} this
method to non-local matrix models defined by a kinetic term (Taylor
coefficient matrix of the two-point function) which is neither
constant nor diagonal. Introducing a cut-off in the measure
$\prod_{m,n} d\phi_{mn}$ of the partition function $Z$, the resulting
effect is undone by adjusting the effective action $L[\phi]$ (and
other terms which are easy to evaluate). If the cut-off function is a
smooth function of the cut-off scale $\Lambda$, the adjustment of
$L[\phi,\Lambda]$ is described by a differential equation,
\begin{align}
 \Lambda \frac{\partial L[\phi,\Lambda]}{\partial \Lambda} &=
  \sum_{m,n,k,l} \frac{1}{2} \Lambda \frac{\partial
    \Delta^K_{nm;lk}(\Lambda)}{\partial \Lambda} \bigg( \frac{\partial
    L[\phi,\Lambda]}{\partial \phi_{mn}}\frac{\partial
    L[\phi,\Lambda]}{\partial \phi_{kl}} - \frac{1}{\mathcal{V}_D}
  \Big[\frac{\partial^2 L[\phi,\Lambda]}{\partial \phi_{mn}\,\partial
    \phi_{kl}}\Big]_\phi \bigg) \;,
\label{polL}
\end{align}
where $\big[F[\phi]\big]_\phi:= F[\phi]-F[0]$ and 
\begin{align}
\Delta^K_{nm;lk}(\Lambda) =
  K[m,n;\Lambda] \Delta_{nm;lk} K[k,l;\Lambda]\;.
\label{GKDK1}  
\end{align}
Here, $K[m,n;\Lambda]$ is the cut-off function which for finite
$\Lambda$ has finite support in $m,n$ and satisfies
$K[m,n;\infty]=1$. By $\mathcal{V}_D$ we denote the volume of an elementary
cell.  

In \cite{Grosse:2003aj} we have derived a power-counting theorem for
$L[\phi,\Lambda]$ by integrating (\ref{polL}) perturbatively between
the initial scale $\Lambda_0$ and the renormalisation scale
$\Lambda_R \ll \Lambda_0$. The power-counting degree is given by
topological data of ribbon graphs and two scaling exponents of the
(summed and differentiated) cut-off propagator. The power-counting
theorem in \cite{Grosse:2003aj} is model independent. The subtraction
of divergences necessary to carry out the limit $\Lambda_0 \to \infty$
has to be worked out model by model.

In this paper we will perform the subtraction of divergences for the
regularised $\phi^4$-model on $\mathbb{R}^2_\theta$. The first step is
to extract from the power-counting theorem \cite{Grosse:2003aj} the
set of relevant and marginal interactions. As we will derive in
Section~\ref{scaling} and Appendix~\ref{appB}, there is an infinite
number of relevant interactions if the regularisation $\Omega$ is not
applied. For $\Omega\neq 0$, which means $\omega<1$, the marginal
interaction is (apart from the initial $\phi^4$-interaction) given by
the planar one-loop two-point function
\begin{align}
  & \parbox{25mm}{\begin{picture}(22,30)
      \put(0,0){\epsfig{file=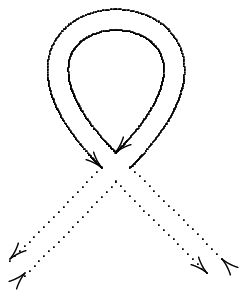}}
      \put(-1,6){\mbox{\scriptsize$m_1$}}
      \put(3,0){\mbox{\scriptsize$n_1$}}
      \put(14,2){\mbox{\scriptsize$m_2$}}
      \put(21,4){\mbox{\scriptsize$n_2$}}
      \put(10,16){\mbox{\scriptsize$l$}}
\end{picture}}
+~~ \parbox{25mm}{\begin{picture}(22,30)
    \put(0,0){\epsfig{file=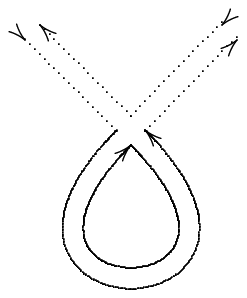}}
    \put(-1,22){\mbox{\scriptsize$n_1$}}
    \put(4,26){\mbox{\scriptsize$m_1$}}
    \put(16,28){\mbox{\scriptsize$n_2$}}
    \put(21,22){\mbox{\scriptsize$m_2$}}
    \put(10,11){\mbox{\scriptsize$l$}}
\end{picture}}
\qquad = \rho_{[m_1]}[\Lambda]\, \delta_{m_1n_2} \delta_{m_2n_1}
+ \rho_{[m_2]}[\Lambda]\, \delta_{m_1n_2} \delta_{m_2n_1}\;.
\label{r1}
\end{align}
For this graph we have to provide boundary conditions at $\Lambda_R$.
The simplicity of the divergent sectors makes the renormalisation very
easy. On the other hand, the simplicity hides the beauty of
renormalisation so that we choose a slightly more general setting to
present the strategy. 

For presentational reasons let us assume that the divergent graphs
have the same structure of external lines as (\ref{r1}) but possibly
an arbitrary number of vertices,
\begin{align}
\parbox{30mm}{\begin{picture}(25,18)
      \put(0,0){\epsfig{file=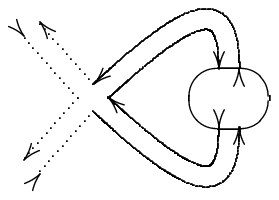}}
      \put(-2,6){\mbox{\scriptsize$m_1$}}
      \put(2,0){\mbox{\scriptsize$n_1$}}
      \put(4,16){\mbox{\scriptsize$m_2$}}
      \put(-2,14){\mbox{\scriptsize$n_2$}}
\end{picture}}  
\qquad = \rho_{[m_2]}[\Lambda]\, \delta_{m_1n_2} \delta_{m_2n_1}\;.
\label{r1g}
\end{align}
In this case the corresponding $\rho_{[m]}$-functions for different indices
$m$ must be expected to be independent, which means that the model would be
determined by an infinite number of free parameters. Since this is not
acceptable, we require according to \cite{Grosse:2003aj} that the parameters
$\rho_{[m]}[\Lambda_R]$ are scaled by the same amount to
$\rho_{[m]}[\Lambda_0]$ (reduction of couplings \cite{Zimmermann:1984sx}).
Expanding $\rho_{[m]}[\Lambda]$ as a formal power series in the coupling
constant $\lambda$, $\rho_{[m]}[\Lambda]=\sum_{V=1}^\infty
\big(\frac{\lambda}{\mu^2}\big)^V \rho^{(V)}_{[m]}[\Lambda]$ and normalising
the renormalised mass $\mu_0$ by $\rho_{[0]}[\Lambda_R]=0$, we thus demand in
general
\begin{align}
\frac{\frac{\mu_0^2}{\mu^2} 
+ \sum_{V'=1}^V \big(\frac{\lambda}{\mu^2}\big)^{V'}
\rho^{(V')}_{[m]}[\Lambda_R]}{\frac{\mu_0^2}{\mu^2} }
\sim \frac{\frac{\mu_0^2}{\mu^2} 
+ \sum_{V'=1}^V \big(\frac{\lambda}{\mu^2}\big)^{V'}
\rho^{(V')}_{[m]}[\Lambda_0]}{\frac{\mu_0^2}{\mu^2} 
+ \sum_{V'=1}^V \big(\frac{\lambda}{\mu^2}\big)^{V'}
\rho^{(V')}_{[0]}[\Lambda_0]}+ \mathcal{O}(\lambda^V)\;.
\label{consistency}
\end{align}
This leads order by order in $\lambda$ to relations $a \sim b$ which
mean $\lim_{\Lambda_0 \to \infty} \frac{a}{b}=1$. 
 
At the initial scale $\Lambda=\Lambda_0$ the effective action thus
reads
\begin{align}
L[\phi,\Lambda_0,\Lambda_0,\omega,\rho^0] 
&= \frac{\lambda}{4!} \sum_{m,n,k,l} \phi_{mn} \phi_{nk} \phi_{kl}
  \phi_{lm} + \frac{1}{2} \sum_{m,n} \rho^0_{[m]} \phi_{mn} \phi_{nm}\;.
\label{ct}
\end{align}
Each summation index runs over $\mathbb{N}$.
The solution of (\ref{polL}) with initial condition (\ref{ct}) will
have a completely different form in terms of $\phi_{mn}$, but the
projection to the same $\phi$-structure as in (\ref{ct}) can still be
defined:
\begin{align}
L[\phi,\Lambda,\Lambda_0,\omega,\rho^0] 
&= \frac{\lambda}{4!} \sum_{m,n,k,l} \phi_{mn} \phi_{nk} \phi_{kl}
  \phi_{lm} + \frac{1}{2} \sum_{m,n} \rho_{[m]}
[\Lambda,\Lambda_0,\omega,\rho^0] \phi_{mn} \phi_{nm}
\nonumber
  \\*
  &+ \text{ different $\phi$-structures}\;.\label{init0}
\end{align}
The marginal part of the four-point function will turn out to be
scale-independent. We identify 
$\rho_{[m]}[\Lambda_0,\Lambda_0,\omega,\rho^0]\equiv \rho^0_{[m]}$. 

At the end we are interested in the limit $\Lambda_0 \to \infty$.
For this purpose we have to admit a $\Lambda_0$-dependence of $\omega$
and $\rho^0_{[m]}$ the determination of which is the art of
renormalisation. For fixed $\Lambda=\Lambda_R$ but
variable $\Lambda_0$ we consider the identity
\begin{align}
&L[\Lambda_R,\Lambda_0',\omega[\Lambda_0'],\rho^0[\Lambda_0']]
-L[\Lambda_R,\Lambda_0'',\omega[\Lambda_0''],\rho^0[\Lambda_0'']]
\nonumber
\\*
& \equiv \int_{\Lambda_0''}^{\Lambda_0'} \frac{d\Lambda_0}{\Lambda_0} \;
\Big(\Lambda_0 \frac{d}{d\Lambda_0}
L[\Lambda_R,\Lambda_0,\omega[\Lambda_0],\rho^0[\Lambda_0]]\Big)
\nonumber
\\*
&=  \int_{\Lambda_0''}^{\Lambda_0'} \frac{d\Lambda_0}{\Lambda_0} \;
\Big(\Lambda_0 \frac{\partial
  L[\Lambda_R,\Lambda_0,\omega,\rho^0]}{\partial\Lambda_0}
+ \Lambda_0 \frac{d \omega}{d \Lambda_0}
\frac{\partial
  L[\Lambda_R,\Lambda_0,\omega,\rho^0]}{\partial\omega}
+ \Lambda_0 \frac{d \rho^0}{d \Lambda_0}
\frac{\partial
  L[\Lambda_R,\Lambda_0,\omega,\rho^0]}{\partial\rho^0}
\Big)\;.
\label{taut}
\end{align}
Here we have omitted for simplicity the dependence of $L$ on $\phi$ as
well as the indices on $\rho^0$.  The model is defined by fixing the
boundary condition for the $\rho$-coefficients at $\Lambda_R$, i.e.\
by keeping $\rho_{[m]}[\Lambda_R,\Lambda_0,\omega,\rho^0]
=\text{constant}$:
\begin{align}
0 &= d \rho_{[m]}[\Lambda_R,\Lambda_0,\omega,\rho^0] 
\nonumber
\\*
&= \frac{\partial \rho_{[m]}[\Lambda_R,\Lambda_0,\omega,\rho^0]}{
\partial \Lambda_0} d\Lambda_0
+ \frac{\partial \rho_{[m]}[\Lambda_R,\Lambda_0,\omega,\rho^0]}{
\partial \omega} d\omega
+ \sum_n \frac{\partial \rho_{[m]}[\Lambda_R,\Lambda_0,\omega,\rho^0]}{
\partial \rho^0_{[n]}} d\rho^0_{[n]}\;.
\end{align}
Assuming that we can invert the matrix 
$\frac{\partial \rho_{[m]}[\Lambda_R,\Lambda_0,\omega,\rho^0]}{
\partial \rho^0_{[n]}}$, which is possible in perturbation theory, we
get
\begin{align}
\frac{d \rho^0_{[n]}}{d \Lambda_0} &= 
- \sum_m \frac{\partial \rho^0_{[n]}}{
\partial \rho_{[m]}[\Lambda_R,\Lambda_0,\omega,\rho^0]} 
\frac{\partial \rho_{[m]}[\Lambda_R,\Lambda_0,\omega,\rho^0]}{\partial
  \Lambda_0} 
\nonumber
\\*
&- \sum_m \frac{\partial \rho^0_{[n]}}{
\partial \rho_{[m]}[\Lambda_R,\Lambda_0,\omega,\rho^0]} 
\frac{\partial \rho_{[m]}[\Lambda_R,\Lambda_0,\omega,\rho^0]}{\partial
  \omega} \frac{d\omega}{d\Lambda_0} \;.
\label{rhoLam}
\end{align}
Inserting (\ref{rhoLam}) into (\ref{taut}) we see that the following
function\footnote{Our function $R$ (for `renormalised') generalises a
  function called $V$ in \cite{Polchinski:1983gv}. We use the symbol
  $R$ in order to avoid confusion with the number $V$ of vertices.
  Below we shall denote the function $B$ of \cite{Polchinski:1983gv}
  by $H$ (for having `holes'), avoiding confusion with the number $B$ of
  boundary components.}  will be important:
\begin{align}
  R[\phi,\Lambda,\Lambda_0,\omega,\rho^0] 
&:= \Lambda_0 \frac{\partial
  L[\phi,\Lambda,\Lambda_0,\omega,\rho^0]}{ \partial \Lambda_0} 
+  \frac{\partial L[\phi,\Lambda,\Lambda_0,\omega,\rho^0]}{\partial \omega}
\Lambda_0 \frac{d \omega}{d \Lambda_0} 
\nonumber
  \\*
  &- \sum_{m,n} \frac{\partial L[\phi,\Lambda,\Lambda_0,\omega,\rho^0]}{
    \partial \rho_{[m]}^0} \frac{\partial \rho_{[m]}^0}{\partial
    \rho_{[n]}[\Lambda,\Lambda_0,\omega,\rho^0]} \Lambda_0 \frac{\partial
    \rho_{[n]}[\Lambda,\Lambda_0,\omega,\rho^0]}{ \partial \Lambda_0}
\nonumber
  \\*
  &- \sum_{m,n} \frac{\partial L[\phi,\Lambda,\Lambda_0,\omega,\rho^0]}{
    \partial \rho_{[m]}^0} \frac{\partial \rho_{[m]}^0}{\partial
    \rho_{[n]}[\Lambda,\Lambda_0,\omega,\rho^0]}  \frac{\partial
    \rho_{[n]}[\Lambda,\Lambda_0,\omega,\rho^0]}{\partial \omega} 
\Lambda_0 \frac{d \omega}{d \Lambda_0}\;.
\label{V}
\end{align}
Now we can rewrite (\ref{taut}) as 
\begin{align}
&L[\phi,\Lambda_R,\Lambda_0',\omega[\Lambda_0'],\rho^0[\Lambda_0']]
- L[\phi,\Lambda_R,\Lambda_0'',\omega[\Lambda_0''],\rho^0[\Lambda_0'']]
\nonumber
\\*
&\qquad\qquad 
= \int_{\Lambda_0''}^{\Lambda_0'} \frac{d\Lambda_0}{\Lambda_0} \;
R[\phi,\Lambda_R,\Lambda_0,\omega[\Lambda_0],\rho^0[\Lambda_0]]\;.
\label{Vlim}
\end{align}
Since $R$ is linear in $L$, the splitting (\ref{init0}) together with
(\ref{V}) leads for all $\Lambda$ to a vanishing projection of $R$ to its
$\rho$-coefficient. In other words, $R$ \emph{projects to the irrelevant part}
of the effective action, which is indispensable for the existence of the limit
$\Lambda_0 \to \infty$ controlled by (\ref{Vlim}). We have to show, however,
that this really eliminates all divergences.

\subsection{Flow equations}

For this purpose we need estimations for $R$. This is achieved by
computing the $\Lambda$-scaling of $R$:
\begin{align}
\Lambda \frac{\partial R}{\partial \Lambda} 
&= \Lambda_0 \frac{\partial}{\partial \Lambda_0}
\Big( \Lambda \frac{\partial  L}{\partial \Lambda}\Big) 
+ \frac{\partial}{\partial \omega}
\Big( \Lambda \frac{\partial  L}{\partial \Lambda}\Big) 
\Lambda_0 \frac{d \omega}{d \Lambda_0} 
\nonumber
\\*
&- \sum_{m,n} \frac{\partial}{\partial \rho_{[m]}^0} 
\Big(\Lambda \frac{\partial L}{\partial \Lambda}\Big) 
\frac{\partial \rho_{[m]}^0}{\partial \rho_{[n]}} 
\Lambda_0 \frac{\partial \rho_{[n]}}{\partial \Lambda_0} 
- \sum_{m,n} \frac{\partial}{\partial \rho_{[m]}^0} 
\Big(\Lambda \frac{\partial L}{\partial \Lambda}\Big) 
\frac{\partial \rho_{[m]}^0}{\partial \rho_{[n]}} 
\frac{\partial \rho_{[n]}}{\partial \omega} 
\Lambda_0 \frac{d \omega}{d \Lambda_0} 
\nonumber
\\*
& + \sum_{m,n,k,l} \frac{\partial L}{\partial \rho_{[m]}^0}
  \frac{\partial \rho_{[m]}^0}{\partial \rho_{[n]}} 
\frac{\partial }{\partial \rho_{[k]}^0} 
\Big(\Lambda \frac{\partial \rho_{[n]}}{\partial \Lambda}\Big) 
\frac{\partial \rho_{[k]}^0}{\partial \rho_{[l]}} 
\Lambda_0 \frac{\partial \rho_{[l]}}{\partial \Lambda_0} 
\nonumber
\\*
&+ \sum_{m,n,k,l} \frac{\partial L}{\partial \rho_{[m]}^0}
  \frac{\partial \rho_{[m]}^0}{\partial \rho_{[n]}} 
\frac{\partial }{\partial \rho_{[k]}^0} 
\Big(\Lambda \frac{\partial \rho_{[n]}}{\partial \Lambda}\Big) 
\frac{\partial \rho_{[k]}^0}{\partial \rho_{[l]}} 
 \frac{\partial \rho_{[l]}}{\partial \omega} 
\Lambda_0\frac{d \omega}{d \Lambda_0} 
\nonumber
\\*
&- \sum_{m,n} \frac{\partial L}{\partial \rho_{[m]}^0} 
\frac{\partial \rho_{[m]}^0}{\partial \rho_{[n]}} 
\Lambda_0 \frac{\partial}{\partial \Lambda_0}
\Big( \Lambda \frac{\partial \rho_{[n]}}{\partial \Lambda}\Big)
- \sum_{m,n} \frac{\partial L}{\partial \rho_{[m]}^0} 
\frac{\partial \rho_{[m]}^0}{\partial \rho_{[n]}} 
\frac{\partial}{\partial \omega} 
\Big( \Lambda \frac{\partial \rho_{[n]}}{\partial \Lambda}\Big)
\Lambda_0 \frac{d \omega}{d \Lambda_0}\;.
\label{VV}
\end{align}
We have omitted the dependencies for simplicity and made use of the
fact that the derivatives with respect to $\Lambda,\Lambda_0,\rho^0,\omega$
commute.  Using (\ref{polL}) we compute the terms on the rhs of
(\ref{VV}):
\begin{align}
  &\Lambda_0 \frac{\partial}{\partial \Lambda_0 }\Big( \Lambda
  \frac{\partial L[\phi,\Lambda,\Lambda_0,\omega,\rho^0]}{\partial \Lambda}
  \Big) \nonumber
  \\*
  &= \sum_{m',n',k',l'} \frac{1}{2} \Lambda \frac{\partial
    \Delta^K_{n'm';l'k'}(\Lambda)}{\partial \Lambda} 
\bigg( 2 \frac{\partial L[\phi,\Lambda,\Lambda_0,\omega,\rho^0]}{
\partial \phi_{m'n'}}  \frac{\partial}{\partial \phi_{k'l'}}
\Big( \Lambda_0 \frac{\partial}{\partial \Lambda_0}
  L[\phi,\Lambda,\Lambda_0,\omega,\rho^0]\Big) \nonumber
  \\*
  &\hspace*{6em} - \frac{1}{\mathcal{V}_2} \Big[\frac{\partial^2}{
    \partial \phi_{m'n'}\,\partial \phi_{k'l'}}\Big( \Lambda_0
  \frac{\partial}{\partial \Lambda_0}
  L[\phi,\Lambda,\Lambda_0,\omega,\rho^0]\Big)\Big]_\phi \bigg) 
%\nonumber
%  \\
%  &
\equiv M\Big[L,\Lambda_0 \frac{\partial L}{\partial \Lambda_0}
  \Big] \;.
\label{V2}
\end{align}
Similarly we have
\begin{align}
  \frac{\partial}{\partial \rho_{[m]}^0}\Big( \Lambda \frac{\partial
    L[\phi,\Lambda,\Lambda_0,\omega,\rho^0]}{\partial \Lambda} \Big) &=
  M\Big[L,\frac{\partial L}{\partial \rho_{[m]}^0}\Big]\;, 
\nonumber
\\*
  \frac{\partial}{\partial \omega}\Big( \Lambda \frac{\partial
    L[\phi,\Lambda,\Lambda_0,\omega,\rho^0]}{\partial \Lambda} \Big) &=
  M\Big[L,\frac{\partial L}{\partial \omega}\Big]\;.
\label{V2a}
\end{align}
In the same way as in (\ref{init0}) we expand $M[L,\,.\,]$ on the rhs
of (\ref{V2}) with respect to the $\phi$-structures,
\begin{align}
M[L,\,.\,]  &= \frac{1}{2} \sum_{m,n} M_{[m]}[L,\,.\,] \phi_{mn} \phi_{nm} 
+ \text{ different $\phi$-structures}\;.
\label{M24}
\end{align}
Because of the $\Lambda$-derivatives there is no analogue of the
initial four-point function. The distinguished expansion coefficients
are due to (\ref{V2}) and (\ref{init0}) identified with
\begin{align}
  M_{[m]}\Big[L, \Lambda_0 \frac{\partial L}{\partial \Lambda_0} \Big] &=
  \Lambda_0 \frac{\partial}{\partial \Lambda_0 }\Big( \Lambda
  \frac{\partial \rho_{[m]}[\Lambda,\Lambda_0,\rho^0]}{\partial \Lambda}
  \Big)\;, 
\nonumber
\\*
 M_{[m]}\Big[L, \frac{\partial L}{\partial \omega} \Big] &=
  \frac{\partial}{\partial \omega}\Big( \Lambda \frac{\partial
    \rho_{[m]}[\Lambda,\Lambda_0,\rho^0]}{\partial \Lambda} \Big)\;,
\nonumber
\\*
 M_{[m]}\Big[L, \frac{\partial L}{\partial \rho^0_{[n]}} \Big] &=
  \frac{\partial}{\partial \rho^0_{[n]}}\Big( \Lambda \frac{\partial
    \rho_{[m]}[\Lambda,\Lambda_0,\rho^0]}{\partial \Lambda} \Big)\;.
\label{MaMa}
\end{align}
Using (\ref{V2}), (\ref{V2a}) and (\ref{MaMa}) as well as the
linearity of $M[L,\,.\,]$ in the second argument we can rewrite
(\ref{VV}) as
\begin{align}
  \Lambda \frac{\partial R}{\partial \Lambda} = M[L,R] - \sum_m
  \frac{\partial L}{\partial \rho_{[m]}} M_{[m]}[L,R]\;,
\label{VVV}
\end{align}
where we have defined
\begin{align}
  \frac{\partial L}{\partial \rho_{[m]}}[\Lambda,\Lambda_0,\omega,\rho^0] :=
  \sum_n \frac{\partial L[\Lambda,\Lambda_0,\omega,\rho^0]}{ \partial
    \rho_{[n]}^0} \frac{\partial \rho_{[n]}^0}{\partial
    \rho_{[m]}[\Lambda,\Lambda_0,\omega,\rho^0]}\;.
\label{Lr}
\end{align}
In the same way as for $R$, the $\Lambda$-scaling of (\ref{Lr}) is computed to
\begin{align}
  \Lambda \frac{\partial}{\partial \Lambda} \Big( \frac{\partial
    L}{\partial \rho_{[m]}} \Big)
  &= M\Big[L, \frac{\partial L}{\partial \rho_{[m]}}\Big] - \sum_{n}
  \frac{\partial L}{\partial \rho_{[n]}} M_{[n]}\Big[L, \frac{\partial
    L}{\partial \rho_{[m]}}\Big]\;.
\label{LLr}
\end{align}

\subsection{Expansion as power series in the coupling constant}

Now we expand the functions just introduced as formal power series in the
coupling constant $\lambda$ and with respect to the number of fields
$\phi$, expressing all dimensionful quantities in terms of the volume
$\mathcal{V}_2$ of the elementary cell:
\begin{align}
  L[\phi,\Lambda%,\Lambda_0,\omega,\rho^0
] &= \lambda \sum_{V =1}^\infty
  \big(\lambda \mathcal{V}_2 \big)^{V-1}
  \sum_{N=2}^\infty \frac{1}{N!} \sum_{m_i,n_i}
  A^{(V)}_{m_1n_1;\dots;m_Nn_N}[\Lambda] \phi_{m_1n_1}\cdots \phi_{m_Nn_N}\;,
\label{Lg}
\\
  R[\phi,\Lambda%,\Lambda_0,\omega,\rho^0
] &= \lambda \sum_{V =1}^\infty
  \big(\lambda \mathcal{V}_2 \big)^{V-1}
  \sum_{N=2}^\infty \frac{1}{N!} \sum_{m_i,n_i}
  R^{(V)}_{m_1n_1;\dots;m_Nn_N}[\Lambda] \phi_{m_1n_1}\cdots \phi_{m_Nn_N}\;,
\label{Rg}
\\
\frac{\partial L}{\partial \rho_{[m]}}[\phi,\Lambda%,\Lambda_0,\omega,\rho^0
] 
&= \sum_{V =0}^\infty
  \big(\lambda \mathcal{V}_2 \big)^{V}
  \sum_{N=2}^\infty \frac{1}{N!} \sum_{m_i,n_i}
  H^{(V)}_{m_1n_1;\dots;m_Nn_N}[\Lambda] \phi_{m_1n_1}\cdots \phi_{m_Nn_N}\;,
\label{Hg}
\end{align}
We have suppressed the additional dependence of $L,R,\frac{\partial
  L}{\partial \rho_{[m]}},A^{(V)},R^{(V)},H^{(V)}$ on
  $\Lambda_0,\omega,\rho^0$.

All functions $A^{(V)},R^{(V)},H^{(V)}$ have mass dimension zero.  The
Polchinski equation (\ref{polL}) as well as its derived equations (\ref{LLr})
and (\ref{VVV}) can now with (\ref{V2}) be written as
\begin{align}
\Lambda \frac{\partial}{\partial \Lambda} 
& A^{(V)}_{m_1n_1;\dots;m_Nn_N}[\Lambda,\Lambda_0,\omega,\rho^0] 
\nonumber
\\*
&= 
\sum_{N_1=2}^N \sum_{V_1=1}^{V-1}
\sum_{m,n,k,l} \frac{1}{2} Q_{nm;lk}(\Lambda)  
A^{(V_1)}_{m_1n_1;\dots;m_{N_1-1}n_{N_1-1};mn}[\Lambda]
 A^{(V-V_1)}_{m_{N_1}n_{N_1};\dots;m_{N}n_{N};kl}[\Lambda]
\nonumber
\\*[-1ex]
&\hspace*{15em} + \Big(\binom{N}{N_1{-}1} -1\Big) \text{ permutations}
\nonumber
\\*
& - \sum_{m,n,k,l} \frac{1}{2} Q_{nm;lk}(\Lambda) 
A^{(V)}_{m_1n_1;\dots;m_{N}n_{N};mn;kl}[\Lambda]\;,
\label{polL2}
\\
\Lambda \frac{\partial}{\partial \Lambda} 
& H^{[\hat{m}](V)}_{m_1n_1;\dots;m_Nn_N}[\Lambda,\Lambda_0,\omega,\rho^0] 
\nonumber
\\*
&= \sum_{N_1=2}^N \sum_{V_1=1}^{V}
\sum_{m,n,k,l} Q_{nm;lk}(\Lambda) 
A^{(V_1)}_{m_1n_1;\dots;m_{N_1-1}n_{N_1-1};mn}[\Lambda]
 H^{[\hat{m}](V-V_1)}_{m_{N_1}n_{N_1};\dots;m_{N}n_{N};kl}[\Lambda]
\nonumber
\\*[-1ex]
&\hspace*{15em} + \Big(\binom{N}{N_1{-}1} -1\Big) \text{ permutations}
\nonumber
\\*
& 
- \sum_{m,n,k,l} \frac{1}{2} Q_{nm;lk}(\Lambda) 
H^{[\hat{m}](V)}_{m_1n_1;\dots;m_{N}n_{N};mn;kl}[\Lambda]
\nonumber
\\*
& 
+ \sum_{\hat{n}} \sum_{V_1=1}^{V}
H^{[\hat{n}](V-V_1)}_{m_1n_1;\dots;m_{N}n_{N}}[\Lambda]
\Big(\sum_{m,n,k,l} \frac{1}{2} Q_{nm;lk}(\Lambda) 
H^{[\hat{m}](V_1)}_{m'n';n'm';mn;kl}[\Lambda]
\Big)_{[\hat{n}]}\;,
\label{polB2}
\\
\Lambda \frac{\partial}{\partial \Lambda} 
& R^{(V)}_{m_1n_1;\dots;m_Nn_N}[\Lambda,\Lambda_0,\omega,\rho^0] 
\nonumber
\\*
&= \sum_{N_1=2}^N \sum_{V_1=1}^{V-1}
\sum_{m,n,k,l} Q_{nm;lk}(\Lambda) 
A^{(V_1)}_{m_1n_1;\dots;m_{N_1-1}n_{N_1-1};mn}[\Lambda]
 R^{(V-V_1)}_{m_{N_1}n_{N_1};\dots;m_{N}n_{N};kl}[\Lambda]
\nonumber
\\*[-1ex]
&\hspace*{15em} + \Big(\binom{N}{N_1{-}1} -1\Big) \text{ permutations}
\nonumber
\\*
& 
- \sum_{m,n,k,l} \frac{1}{2} Q_{nm;lk}(\Lambda) 
R^{(V)}_{m_1n_1;\dots;m_{N}n_{N};mn;kl}[\Lambda]
\nonumber
\\*
& 
+ \sum_{\hat{n}} \sum_{V_1=1}^{V}
H^{[\hat{n}](V-V_1)}_{m_1n_1;\dots;m_{N}n_{N}}[\Lambda]
\Big(\sum_{m,n,k,l} \frac{1}{2} Q_{nm;lk}(\Lambda) 
R^{(V_1)}_{m'n';n'm';mn;kl}[\Lambda]
\Big)_{[\hat{n}]}\;,
\label{polV2}
\end{align}
with
\begin{align}
  Q_{nm;lk}(\Lambda) := \frac{1}{\mathcal{V}_2} \Lambda
  \frac{\partial \Delta^K_{nm;lk}(\Lambda)}{\partial \Lambda}\;.
\label{Q}
\end{align}
Note that the projection $(~)_{[m]}$ to the $\rho_{[m]}$-coefficients
in (\ref{polB2}) and (\ref{polV2}) are due to (\ref{r1g}) non-zero on
the 1PI functions only.

\section{Renormalisation of the $\phi^4$-model}

\subsection{Scaling of the cut-off propagator}
\label{scaling}

We have
$\mathcal{V}_2=2\pi\theta_1=\frac{8\pi}{(1+\sqrt{\omega})\mu^2}$.
We choose the smooth cut-off function
\begin{align}
K(m,n;\Lambda) &=  K\Big(\frac{m \mu^2}{\Lambda^2}\Big) 
K\Big(\frac{n \mu^2}{\Lambda^2}\Big) \;, \qquad \text{where}
\nonumber
\\*
K(x) &\in  C^\infty(\mathbb{R}^+) \text{ is monotonous with }
K(x) = \left\{ \begin{array}{ll}  1 \qquad& \text{for } x\leq 1\;,\\
    0 \qquad& \text{for } x\geq 2\;.
\end{array}\right.
\label{cut-off}
\end{align}
This choice satisfies the dimensional normalisation 
\begin{align}
\sum_{m} \text{sign} \Big(
\max_{n,l} \big|K(m,n;\Lambda) K(l{+}n{-}m,l;\Lambda)\big| \Big)
\leq  \sum_{m=0}^{\frac{2\Lambda^2}{\mu^2}-1} 1 
= 2 \Big(\frac{\Lambda}{\mu}\Big)^2
\label{volumefactor}
\end{align}
of a two-dimensional model \cite{Grosse:2003aj}. We obtain with (\ref{GKDK1}) 
\begin{align}
\Lambda \frac{\partial \Delta_{nm;lk}^K(\Lambda)}{\partial \Lambda}
  &= - \sum_{j\in\{m,n,k,l\}} \frac{2 j \mu^2}{\Lambda^2} 
K'\Big(\frac{j \mu^2}{\Lambda^2}\Big) \prod_{i \in \{m,n,k,l\}\setminus
  \{j\}} K\Big(\frac{i\mu^2}{\Lambda^2}\Big) \Delta_{nm;lk}\;.
\label{DK1}
\end{align}
Since $\mathrm{supp}\, K'(x)=[1,2]$ and $\mathrm{supp}\, K(y)=[0,2]$,
(\ref{DK1}) is non-zero only if the condition
\begin{align}
  \frac{\Lambda^2}{\mu^2} \leq \max(m,n,k,l) &\leq
  \frac{2\Lambda^2}{\mu^2} 
\label{Qc}
\end{align}
is satisfied. Note that due to (\ref{pmax}) and (\ref{mu}) this
also corresponds to a momentum cut-off $p_{\text{max}} \approx \sqrt{8}
\Lambda$. We compute in Appendix~\ref{appB} the
$\Lambda$-dependence of the maximised propagator for selected values of
$\mu_0$ and $\omega$, which is extremely well reproduced by
(\ref{Form1}). We thus obtain for the maximum of (\ref{Q}) 
\begin{align}
\big|Q_{nm;lk}(\Lambda)\big| & \leq
  \frac{(1{+}\sqrt{\omega})\mu^2}{8\pi} \,(16 \max_x
  |K'(x)|)\,\big|\Delta^{\mathcal{C}}_{nm;lk}\big|_{\mathcal{C}
    =\frac{\Lambda^2}{\mu^2}}
\nonumber  
\\*
& \leq \left\{ \begin{array}{ll} \displaystyle 
C_0 \frac{\mu^2}{(1{-}\omega)^{\frac{1}{2}} \Lambda^2}\,
      \delta_{m+k,n+l}\qquad & \text{for } \omega<1\;,
      \\[2ex]
      \displaystyle C_0 \frac{\mu^2}{\mu_0 \Lambda}\,
      \delta_{m+k,n+l} \qquad & \text{for } \omega=1\;,
\end{array}\right.
\label{est0}
\end{align}
where $C_0=0.78\,C_0'\, \max_x |K'(x)|$.  The constant $C_0'
\gtrapprox 1$ corrects the fact that (\ref{Form1}) holds
asymptotically only.  Next, from (\ref{Form2}) we obtain
\begin{align}
 \max_n \sum_{k} \max_{m,l}
  \big|Q_{nm;lk}(\Lambda)\big| & \leq
  \frac{(1{+}\sqrt{\omega})\mu^2}{8\pi} \,(16 \max_x |K'(x)|) 
  \max_n \sum_{k} \max_{m,l}
  \big|\Delta_{nm;lk}^{\mathcal{C}}
  \big|_{\mathcal{C}=\frac{\Lambda^2}{\mu^2}} \nonumber
  \\*
  & \leq \left\{ \begin{array}{ll} \displaystyle 
C_1 \frac{\mu^2}{(1{-}\omega) \Lambda^2}\qquad & \text{for }
      \omega<1\;,
      \\[2ex]
      \displaystyle C_1 \frac{\mu^2}{\mu_0^2} \qquad & \text{for }
      \omega=1\;,
    \end{array}\right.
    \label{est1}
\end{align}
where $C_1=1.28\,C_1'\, \max_x |K'(x)|$. We conclude from
\cite{Grosse:2003aj} that the scaling exponents of the propagator are
given by
\begin{align}
\delta_0&=\delta_1=2  \qquad \text{for } \omega<1\;, &
\delta_0&=1\;,~~\delta_1=0  \qquad \text{for }\omega=1\;.
\end{align}
We thus have a regular model for $\omega<1$ and an anomalous (and not
renormalisable) model for $\omega=1$. We also need the product of
(\ref{est0}) with (\ref{volumefactor}):
\begin{align}
&\max_{m,n,k,l}  \big|Q_{nm;lk}(\Lambda)\big|\;
\sum_{m'} \text{sign} \Big(
\max_{n',l'} |K(m',n';\Lambda) K(l'{+}n'{-}m',l';\Lambda)| \Big)
\nonumber
\\*
&\qquad \leq \left\{ \begin{array}{ll} \displaystyle 
2 C_0 \frac{1}{(1{-}\omega)^{\frac{1}{2}}} \qquad & \text{for } \omega<1\;,
      \\[2ex]
\displaystyle 2 C_0 \frac{\Lambda}{\mu_0} \qquad & \text{for } \omega=1\;.
\end{array}\right.
\label{est2}
\end{align}

\subsection{Verification of the consistency condition}

We first have to verify the consistency condition (\ref{consistency}), 
which in the present case simplifies considerably. Since the expansion
stops at first order in the coupling constant,
$\rho^{(V)}_{[m]}\equiv 0$ for $V>1$, we get the condition 
\begin{align}
\rho_{[0]}[\Lambda_0]\sim
\rho_{[m]}[\Lambda_0]-\rho_{[m]}[\Lambda_R]\;.
\end{align}
The initial value $\rho_{[m]}[\Lambda_R]$ drops out, and according to
(\ref{r1}) we have to verify
\begin{align}
1 = \lim_{\Lambda_0 \to \infty}  
\frac{\displaystyle \int_{\Lambda_R}^{\Lambda_0}
  \frac{d\Lambda}{\Lambda} \sum_n Q_{nm;mn}(\Lambda)}{
\displaystyle \int_{\Lambda_R}^{\Lambda_0}
  \frac{d\Lambda}{\Lambda} \sum_n Q_{n0;0n}(\Lambda)}
\equiv \lim_{\Lambda_0 \to \infty}  
\frac{\sum_n
  (\Delta^K_{nm;mn}(\Lambda_0)-\Delta^K_{nm;mn}(\Lambda_R))}{
\sum_n (\Delta^K_{n0;0n}(\Lambda_0)-\Delta^K_{n0;0n}(\Lambda_R))}\;,
\label{lim1}
\end{align}
where we have used (\ref{Q}). Let $\Lambda_m \ll \Lambda_0$ be the minimal
scale such that for $n\geq \frac{\Lambda_m^2}{\mu^2}$ the value of the
propagator $\Delta_{nm;mn}$ lies in the interval formed by the two asymptotics
of Figure~\ref{fig3}. We have $\Lambda_m^2 \approx 2 C_m m \mu^2$ where $C_m$
is of order $1$. Then we have with (\ref{asDelta})
\begin{align}
\sum_{n=0}^{\frac{\Lambda_m^2}{\mu^2}-1} \Delta_{nm;mn}
+ \sum_{n=\frac{\Lambda_m^2}{\mu^2}}^{\frac{\Lambda_0^2}{\mu^2}-1}
\frac{1}{\mu^2(n-\frac{9\omega-5}{4}m+5)}
&< \sum_n   \Delta^K_{nm;mn}(\Lambda_0) 
\nonumber
\\*[-2ex]
&< \sum_{n=0}^{\frac{\Lambda_m^2}{\mu^2}-1} \Delta_{nm;mn}
+ \sum_{n=\frac{\Lambda_m^2}{\mu^2}}^{\frac{2\Lambda_0^2}{\mu^2}-1}
\frac{1}{\mu^2(n-\frac{9\omega-5}{4}m-2)}\;.
\end{align}
This shows that $ \sum_n \Delta^K_{nm;mn}(\Lambda_0)$ is
logarithmically divergent for $\Lambda_0\to \infty$ and that
(\ref{lim1}) holds independently of the finite quantities $\sum_n
\Delta^K_{nm;mn}(\Lambda_R)$ and
$\sum_{n=0}^{\frac{\Lambda_m^2}{\mu^2}-1} \Delta_{nm;mn}$ and
independently of the cut-off function (\ref{cut-off}).

\subsection{Estimations for the interaction coefficients}

According to \cite{Grosse:2003aj} the Polchinski equation
(\ref{polL2}) is solved by ribbon graphs characterised by the number
$V$ of vertices, the number $V^e$ of external vertices, the number $B$
of boundary components, the genus $\tilde{g}$ and the segmentation
index $\iota$. We also recall that it is necessary to sum over indices
of the external legs of ribbon graphs.  There are $s\leq V^e+ \iota
- 1$ summations over different outgoing indices where the
corresponding incoming index of the trajectories are kept fixed. We
write symbolically $\sum_{\mathcal{E}^s}$ for the index summation. 

We can now quote directly the power-counting theorem proven in
\cite{Grosse:2003aj}, inserting (\ref{est0}), (\ref{est1}) and (\ref{est2}):
\begin{lem} 
\label{lemA}
The homogeneous parts $A^{(V,V^e,B,\tilde{g},\iota)}_{
  m_1n_1;\dots;m_Nn_N}$ of the coefficients of the effective
action describing a regularised $\phi^4$-theory on
$\mathbb{R}^2_\theta$ in the matrix base are for $2 \leq N\leq 2V{+}2$
and $\sum_{i=1}^N (m_i{-}n_i)=0$ bounded by
\begin{align}
&\sum_{\mathcal{E}^{s}} \big|A^{(V,V^e,B,\tilde{g},\iota)}_{
    m_1n_1;\dots;m_Nn_N}[\Lambda,\Lambda_0,\omega,\rho_0] \big| 
\nonumber
  \\*
  &\quad 
\leq \Big(\frac{\Lambda^2}{\mu^2}\Big)^{2-V-B-2\tilde{g}}
  \Big(\frac{1}{\sqrt{1{-}\omega}}\Big)^{3V-\frac{N}{2}-1 
+B+2\tilde{g}-V^e-\iota+s}
  \,P^{2V-\frac{N}{2}}\Big[\ln \frac{\Lambda_0}{\Lambda_R}\Big]\;,
\label{ANnorm1}
\end{align}
where $P^q[X]$ denotes a polynomial in $X$ up to degree $q$.  We have
$A^{(V,V^e,B,\tilde{g},\iota)}_{ m_1n_1;\dots;m_Nn_N}\equiv 0$ for
$N>2V{+}2$ or $\sum_{i=1}^N (m_i{-}n_i)\neq 0$.  \hfill $\square$
\end{lem}
The choice of the boundary conditions is at the same time determined
by (\ref{ANnorm1}) and required to prove (\ref{ANnorm1}). We notice
that the marginal interaction coefficients are those with $V= B = 1$
(and $\tilde{g}= 0$, but this holds automatically for $V = 1$).  We
can impose the boundary conditions for
$A^{(1,1,1,0,0)}_{m_1n_1;\dots;m_4n_4}$ at $\Lambda_0$ whereas for
$A^{(1,1,1,0,0)}_{m_1n_1;m_2n_2}$ the limit $\Lambda_0 \to \infty$
later on requires to choose the boundary condition at $\Lambda_R$. We
thus demand
\begin{align}
A^{(1,1,1,0,0)}_{m_1n_1;\dots;m_4n_4}[\Lambda_0,\Lambda_0,\omega,\rho^0] 
&= \frac{1}{6} \Big(\delta_{n_1m_2} \delta_{n_2m_3}\delta_{n_3m_4}
\delta_{n_4m_1} + 5 \text{ permutations } \Big)\;,
\nonumber
\\*
A^{(1,1,1,0,0)}_{m_1n_1;m_2n_2}[\Lambda_R,\Lambda_0,\omega,\rho^0] 
& \equiv (\rho_{[m_1]}+ \rho_{[m_2]})[\Lambda_R,\Lambda_0,\omega,\rho^0] 
\delta_{m_1n_2} \delta_{m_2n_1} = 0 \;,
\nonumber
\\*
A^{(V,V^e,B,\tilde{g},\iota)}_{m_1n_1;\dots;m_Nn_N}
[\Lambda_0,\Lambda_0,\omega,\rho^0] &= 0\qquad \text{for all } V+B>2\;.
\label{initcond}
\end{align}

We remark that for $\omega = 1$ and an optimal choice of the boundary
conditions for $A^{(V;V^e,B,\tilde{g},\iota)}_{m_1n_1;\dots;m_Nn_N}$ in
agreement with \cite{Grosse:2003aj} we would get
\begin{align}
&\sum_{\mathcal{E}^{s}} \big|A^{(V,V^e,B,\tilde{g},\iota)}_{
    m_1n_1;\dots;m_Nn_N}[\Lambda,\Lambda_0,1,\rho^0] \big| 
\nonumber
  \\*
&\quad 
\leq \Big(\frac{\Lambda}{\mu}\Big)^{V-\frac{N}{2}+3-B-2\tilde{g}
-V^e-\iota+s}
  \Big(\frac{\mu}{\mu_0}\Big)^{3V-\frac{N}{2}-1 
+B+2\tilde{g}-V^e-\iota+s}
  \,P^{2V-\frac{N}{2}}\Big[\ln \frac{\Lambda_0}{\Lambda_R}\Big]\;.
\label{ANnorm2}
\end{align}
There would be an infinite number of relevant interaction coefficients,
which means that the model is not renormalisable when keeping $\omega =
1$.

For the limit $\Lambda_0 \to \infty$ of the theory we are interested
in the functions $R^{(V)}_{m_1n_1;\dots;m_Nn_N}$, see (\ref{Vlim}).
The $R^{(V)}_{m_1n_1;\dots;m_Nn_N}$ are the solution of the
differential equation (\ref{polV2}) given again by ribbon graphs. These
graphs are identical to the graphs representing the $A$-functions.
The differential equation (\ref{polV2}) actually simplifies in the
model under consideration because for $\omega < 1$ the projection
$(~)_{[\hat{n}]}$ is of at most first order in the coupling constant.
This means that
\begin{align}
\Big(\sum_{m,n,k,l} \frac{1}{2} Q_{nm;lk}(\Lambda) 
R^{(V_1)}_{m'n';n'm';mn;kl}[\Lambda]
\Big)_{[\hat{n}]}=0 \qquad \text{unless } V_1=1
\end{align}
in the last line of (\ref{polV2}). However, the rhs of (\ref{polV2})
for $V = 1$ and $N = 4$ is identically zero, because
$R^{(1)}_{m_1n_1;\dots;m_6n_6}=0$ by graphical reasons and
$H^{(0)}_{m_1n_1;\dots;m_4n_4}=0$ due to the fact that
$A^{(1)}_{m_1n_1;\dots;m_4n_4}[\Lambda]=
A^{(1)}_{m_1n_1;\dots;m_4n_4}[\Lambda_0]$ is independent of
$\rho^0_{[m]}$, see (\ref{Lr}) and (\ref{initcond}). We thus obtain
\begin{align}
&R^{(1)}_{m_1n_1;\dots;m_4n_4}[\Lambda,\Lambda_0,\omega,\rho^0]
=R^{(1)}_{m_1n_1;\dots;m_4n_4}[\Lambda_0,\Lambda_0,\omega,\rho^0]
\nonumber
\\*
&=\Lambda_0 \frac{\partial}{\partial \Lambda_0} 
A^{(1)}_{m_1n_1;\dots;m_4n_4}[\Lambda,\Lambda_0,\omega,\rho^0]
\Big|_{\Lambda=\Lambda_0} 
+ \frac{\partial A^{(1)}_{m_1n_1;\dots;m_4n_4}[\Lambda_0,\Lambda_0,
\omega,\rho^0]}{\partial \omega}
\Lambda_0 \frac{d \omega}{d \Lambda_0} 
\nonumber
\\*
&- \sum_n H^{[n](0)}_{m_1n_1;\dots;m_4n_4}[\Lambda_0,\Lambda_0,
\omega,\rho^0] \Big( \Lambda_0 \frac{\partial}{\partial \Lambda_0} 
\rho_{[n]}[\Lambda,\Lambda_0,\omega,\rho^0]
\Big|_{\Lambda=\Lambda_0} \!\!
+ \frac{\partial \rho_{[n]}[\Lambda,\Lambda_0,\omega,\rho^0]}{\partial \omega}
\Lambda_0 \frac{d \omega}{d \Lambda_0} \Big) 
\nonumber
\\*
&=0 \;.
\label{R4}
\end{align}
We have $R^{(1,1,1,0,0)}_{m_1n_1;m_2n_2}=0$ by definition (\ref{V}).
The conclusion is that (\ref{polV2}) simplifies to
\begin{align}
\Lambda \frac{\partial}{\partial \Lambda} 
& R^{(V)}_{m_1n_1;\dots;m_Nn_N}[\Lambda] 
\nonumber
\\*
&= \sum_{N_1=2}^N \sum_{V_1=1}^{V-1}
\sum_{m,n,k,l} Q_{nm;lk}(\Lambda) 
A^{(V_1)}_{m_1n_1;\dots;m_{N_1-1}n_{N_1-1};mn}[\Lambda]
 R^{(V-V_1)}_{m_{N_1}n_{N_1};\dots;m_{N}n_{N};kl}[\Lambda]
\nonumber
\\*[-1ex]
&\hspace*{15em} + \Big(\binom{N}{N_1{-}1} -1\Big) \text{ permutations}
\nonumber
\\*
& 
- \sum_{m,n,k,l} \frac{1}{2} Q_{nm;lk}(\Lambda) 
R^{(V)}_{m_1n_1;\dots;m_{N}n_{N};mn;kl}[\Lambda]\;.
\label{polV2beta}
\end{align}
Hence, we do not have to evaluate the $H$-functions for $\omega < 1$.

\begin{lem} 
\label{lemV}
The homogeneous parts $R^{(V,V^e,B,\tilde{g},\iota)}_{
  m_1n_1;\dots;m_Nn_N}$ of the coefficients of the
$\Lambda_0$-varied effective action describing a regularised
$\phi^4$-theory on $\mathbb{R}^2_\theta$ in the matrix base are for $2
\leq N\leq 2V{+}2$ and $\sum_{i=1}^N (m_i{-}n_i)=0$ bounded by
\begin{align}
&\sum_{\mathcal{E}^{s}} \big|R^{(V,V^e,B,\tilde{g},\iota)}_{
    m_1n_1;\dots;m_Nn_N}[\Lambda,\Lambda_0,\omega,\rho^0] \big|
\nonumber
 \\*
 &\quad 
\leq \frac{\Lambda^2}{\Lambda_0^2} 
\Big(\frac{\Lambda^2}{\mu^2}\Big)^{2-V-B-2\tilde{g}}
  \Big(\frac{1}{\sqrt{1{-}\omega}}\Big)^{3V-\frac{N}{2}-1 
+B+2\tilde{g}-V^e-\iota+s}
  \,P^{2V-\frac{N}{2}}\Big[\ln \frac{\Lambda_0}{\Lambda_R}\Big]\;,
\label{VNnorm1}
\end{align}
for $V+B>2$. We have $R^{(V,V^e,B,\tilde{g},\iota)}_{
  m_1n_1;\dots;m_Nn_N}\equiv 0$ for $N>2V{+}2$, for $V+B=2$
or for $\sum_{i=1}^N (m_i{-}n_i)\neq 0$.
\end{lem}
\textit{Proof.} We first derive the initial condition. From
(\ref{initcond}) we learn that for $V+B>2$ we have
$A^{(V,V^e,B,\tilde{g},\iota)}_{
  m_1n_1;\dots;m_Nn_N}[\Lambda_0,\Lambda_0,\omega,\rho^0]\equiv 0$
\emph{independent of} $\Lambda_0,\omega,\rho^0$:
\begin{align}
0&=  \Lambda_0 \frac{\partial}{\partial \Lambda_0} 
A^{(V,V^e,B,\tilde{g},\iota)}_{m_1n_1;\dots;m_Nn_N}
[\Lambda_0,\Lambda_0,\omega,\rho^0]
\nonumber
\\*
& =  \frac{\partial}{\partial \omega} 
A^{(V,V^e,B,\tilde{g},\iota)}_{m_1n_1;\dots;m_Nn_N}
[\Lambda_0,\Lambda_0,\omega,\rho^0]
=  \frac{\partial}{\partial \rho^0} 
A^{(V,V^e,B,\tilde{g},\iota)}_{m_1n_1;\dots;m_Nn_N}
[\Lambda_0,\Lambda_0,\omega,\rho^0]\;,
\label{initV}
\end{align}
for $V+B>2$. The first line has to be considered with care:
\begin{align}
0 =\Lambda_0 \frac{\partial}{\partial \Lambda_0}
A^{(V,V^e,B,\tilde{g},\iota)}_{m_1n_1;\dots;m_Nn_N}
[\Lambda_0,\Lambda_0,\omega,\rho^0]
&\equiv  
\Lambda_0 \frac{\partial}{\partial \Lambda_0}
A^{(V,V^e,B,\tilde{g},\iota)}_{m_1n_1;\dots;m_Nn_N}
[\Lambda,\Lambda_0,\omega,\rho^0]\Big|_{\Lambda=\Lambda_0} 
\nonumber
\\*
&+ \Lambda \frac{\partial}{\partial \Lambda}
A^{(V,V^e,B,\tilde{g},\iota)}_{m_1n_1;\dots;m_Nn_N}
[\Lambda,\Lambda_0,\omega,\rho^0]\Big|_{\Lambda=\Lambda_0} \;.
\label{LL0}
\end{align}
Inserting (\ref{ANnorm1}) into (\ref{LL0}) and further into
(\ref{initV}) we obtain the initial condition for the functions $R$ defined in
(\ref{V}) as
\begin{align}
&\sum_{\mathcal{E}^{s}} \big|R^{(V,V^e,B,\tilde{g},\iota)}_{
    m_1n_1;\dots;m_Nn_N}[\Lambda_0,\Lambda_0,\omega,\rho^0] \big|
\nonumber
\\*
&\qquad \leq 
\Big(\frac{\Lambda_0^2}{\mu^2}\Big)^{2-V-B-2\tilde{g}}
  \Big(\frac{1}{\sqrt{1{-}\omega}}\Big)^{3V-\frac{N}{2}-1 
+B+2\tilde{g}-V^e-\iota+s}
  \,P^{2V-\frac{N}{2}}\Big[\ln \frac{\Lambda_0}{\Lambda_R}\Big]\;.
\label{VNnorm2}
\end{align}

Because of (\ref{R4}) we obtain from (\ref{polV2beta}) for the simplest
non-vanishing $R$-functions
\begin{align}
R^{(1,1,2,0,1)}_{m_1n_1;m_2n_2}[\Lambda]&=  
R^{(1,1,2,0,1)}_{m_1n_1;m_2n_2}[\Lambda_0]\;, &
R^{(2,2,1,0,0)}_{m_1n_1;\dots;m_6n_6}[\Lambda]&=  
R^{(2,2,1,0,0)}_{m_1n_1;\dots;m_6n_6}[\Lambda_0]\;,
\label{R6}
\end{align}
which due to (\ref{VNnorm2}) are in agreement with (\ref{VNnorm1}).
Since (\ref{polV2beta}) is a linear differential equation, the factor
$\frac{\Lambda^2}{\Lambda_0^2}$ first appearing in (\ref{R6}) survives
to more complicated graphs.  Indeed, the only difference between
(\ref{VNnorm1}) and (\ref{ANnorm1}) is the factor
$\frac{\Lambda^2}{\Lambda_0^2}$, and the structure of the rhs of the
differential equation (\ref{polV2beta}) and is the same as for
(\ref{polL2}).  We can thus repeat the evaluation of the Polchinski
equation (\ref{polL2}) performed in \cite{Grosse:2003aj} for the
similar differential equation (\ref{polV2beta}). We find immediately
by induction that the rhs of (\ref{polV2beta}) is bounded by
(\ref{VNnorm1}) with the degree of the polynomial in
$\ln\frac{\Lambda_0}{\Lambda_R}$ reduced by $1$. This leads to
\begin{align}
&
\sum_{\mathcal{E}^{s}} \big|R^{(V,V^e,B,\tilde{g},\iota)}_{
    m_1n_1;\dots;m_Nn_N}[\Lambda,\Lambda_0,\omega,\rho^0] \big| 
\nonumber
\\*
& \quad \leq 
\sum_{\mathcal{E}^{s}} \big|R^{(V,V^e,B,\tilde{g},\iota)}_{
    m_1n_1;\dots;m_Nn_N}[\Lambda_0,\Lambda_0,\omega,\rho^0] \big| 
\nonumber
\\*
& \quad + \int_{\Lambda}^{\Lambda_0} \frac{d\Lambda'}{\Lambda'} 
\frac{\Lambda^{\prime 2}}{\Lambda_0^2} 
\Big(\frac{\Lambda^{\prime 2}}{\mu^2}\Big)^{2-V-B-2\tilde{g}}
  \Big(\frac{1}{\sqrt{1{-}\omega}}\Big)^{3V-\frac{N}{2}-1 
+B+2\tilde{g}-V^e-\iota+s}
  \,P^{2V-\frac{N}{2}-1}\Big[\ln \frac{\Lambda_0}{\Lambda_R}\Big]\;.
\end{align}
Since $V+B>2$ the integral is bounded by (\ref{VNnorm1}). \hfill $\square$
\bigskip

We have convinced ourselves that it is crucial to keep $\omega<1$.  We are,
however, interested in the standard $\phi^4$-model given by $\Omega=0$ and
thus $\omega=(\frac{1-\Omega^2}{1+\Omega^2})^2=1$. \emph{This model can be
  achieved in the limit.} For this purpose we have to find a dependence
$\omega[\Lambda_0]$ with $\lim_{\Lambda_0\to\infty} \omega[\Lambda_0]=1$ which
additionally leads to convergence of (\ref{Vlim}). One choice which meets the
criteria is 
\begin{align}
\omega[\Lambda_0] &=1 - \Big(1+\ln \frac{\Lambda_0}{\Lambda_R}\Big)^{-2}\;, &
\Lambda_0 \frac{d \omega[\Lambda_0]}{d\Lambda_0} &= 
2\Big(1+\ln \frac{\Lambda_0}{\Lambda_R}\Big)^{-3} \equiv
2\big(1{-}\omega[\Lambda_0]\big)^{\frac{3}{2}} \;.
\label{bL0}
\end{align}

\begin{thm}
  The $\phi^4$-model on $\mathbb{R}^2_\theta$ is (order by order in the
  coupling constant) renormalisable in the
  matrix base by adjusting the
  coefficients $\rho^0_{[m]}[\Lambda_0]$ of the initial interaction to
  give $A^{(1;1;1;0;0)}_{m_1n_1;m_2n_2}[\Lambda_R] = 0$ and by
  performing the limit $\Lambda_0 \to \infty$ along the path of
  regulated models characterised by $\omega[\Lambda_0] = 1 - (1+\ln
  \frac{\Lambda_0}{\Lambda_R})^{-2}$.  The limit
  $A^{(V,V^e,B,\tilde{g},\iota)}_{m_1n_1;\dots;m_Nn_N}[\Lambda_R,\infty]
  :=\lim_{\Lambda_0 \to \infty}
  A^{(V,V^e,B,\tilde{g},\iota)}_{m_1n_1;\dots;m_Nn_N}
  [\Lambda_R,\Lambda_0,\omega[\Lambda_0],\rho^0[\Lambda_0]]$ of the
  expansion coefficients of the effective action
  $L[\phi,\Lambda_R,\Lambda_0,\omega[\Lambda_0],\rho^0[\Lambda_0]]$,
  see (\ref{Lg}), exists and satisfies
\begin{align}
  & \Big| \lambda \big(\lambda \mathcal{V}_2\big)^{V-1}
  A^{(V,V^e,B,\tilde{g},\iota)}_{m_1n_1;\dots;m_Nn_N}[\Lambda_R,\infty] -
  \big(\lambda \mathcal{V}_2\big)^{V-1}
  A^{(V,V^e,B,\tilde{g},\iota)}_{m_1n_1;\dots;m_Nn_N}
[\Lambda_R,\Lambda_0,\omega,\rho^0]
  \Big|_{\omega = 1 - (1+\ln \frac{\Lambda_0}{\Lambda_R})^{-2}} \nonumber
  \\
  &\qquad \leq \frac{\Lambda_R^4}{\Lambda_0^2}
  \Big(\frac{\lambda}{\Lambda_R^2}\Big)^V \Big(\frac{\mu^2 (1+\ln
    \frac{\Lambda_0}{\Lambda_R})}{\Lambda_R^2} \Big)^{B+2\tilde{g}-1}
  \,P^{5V-N-V^e-\iota}\Big[\ln \frac{\Lambda_0}{\Lambda_R}\Big]\;.
\label{limA}
\end{align}
\end{thm}
\textit{Proof.} The question is whether
$L[\phi,\Lambda_R,\Lambda_0,\omega[\Lambda_0],\rho^0[\Lambda_0]]$ converges
to a finite limit when $\Lambda_0 \to \infty$.  The existence of the
limit and its property (\ref{limA}) follow from inserting
(\ref{VNnorm1}) and (\ref{bL0}) into (\ref{Vlim}) and Cauchy's
criterion. Note that $\int \frac{dx}{x^3}\,P^q[\ln x] = \frac{1}{x^2} 
P^{\prime q}[\ln x]$. \hfill $\square$% 
\bigskip

It seems that we can additionally achieve a commutative theory
$\theta_1 = \frac{2}{\mu^2} \to 0$ in the limit $\Lambda_0 \to \infty$
by choosing e.g. $\mu^2 = \Lambda_R^2 \sqrt{1+\ln
  \frac{\Lambda_0}{\Lambda_R}}$. (We need
$4\Omega/\theta_1=\mu^2\sqrt{1-\omega} \to 0$.) However, this limit is 
degenerate because due to (\ref{Qc}) all indices are frozen to zero. A
different reference scale than $\mu$ would help, but we need precisely
the choice (\ref{Qc}) in order to get the correct momentum cut-off from
(\ref{pmax}). There is additional work necessary to get the
commutative limit from (\ref{limA}).

\section{Conclusion}

Using the adapted Wilson-Polchinski approach developed in
\cite{Grosse:2003aj} we have proven that the real $\phi^4$-model on
$\mathbb{R}^2_\theta$ is perturbatively renormalisable when formulated
in the matrix base. It was crucial to define the model at the initial
scale $\Lambda_0$ by the $\phi^4$-action supplemented by a harmonic
oscillator potential. The renormalisation is achieved by a suitable
$\Lambda_0$-dependence of the bare mass and the oscillator frequency.
This shows that the limit $\Lambda_0 \to \infty$ of our model is
different from the subtraction of divergences arising in the
na\"{\i}ve Feynman graph approach in momentum space. Whereas the
treatment of the oscillator potential is easy in the matrix base, a
similar procedure in momentum space will face enormous difficulties.
In contrast to the Feynman graph approach, our renormalised Green's
functions are bounded. 

First calculations of the asymptotic behaviour
of the propagator in the four-dimensional case suggest that by the
same regulator method it will be possible to renormalise the
$\phi^4$-model on $\mathbb{R}^4_\theta$ \cite{gw3}.

\section*{Acknowledgement}

We would like to thank the Erwin Schr\"odinger Institute in Vienna and
the Max-Planck-Institute for Mathematics in the Sciences in Leipzig
for hospitality during several mutual visits as well as for the
financing of these invitations.

\begin{appendix}

\section{The matrix basis of $\mathbb{R}^2_\theta$}
\label{appA}

The following is copied from \cite{Gracia-Bondia:1987kw}, adapted to
our notation. The Gaussian
\begin{align}
  f_0(x) &= 2 \mathrm{e}^{-\frac{1}{\theta_1}(x_1^2+x_2^2)}\;,
\end{align}
with $\theta_1\equiv \theta^{12}=-\theta^{21}>0$, is an idempotent,
\begin{align}
  (f_0 \star f_0)(x) &= 4 \int d^2y\int \frac{d^2k}{(2\pi)^2} \,
  \mathrm{e}^{-\frac{1}{\theta_1}(2 x^2 +y^2+ 2x\cdot y + x \cdot \theta
    \cdot k + \frac{1}{4} \theta_1^2 k^2) + \mathrm{i} k\cdot y}
= f_0(x)\;.
\label{ff}
\end{align}
We consider creation and annihilation operators
\begin{align}
  a &= \frac{1}{\sqrt{2}}(x_1+\mathrm{i} x_2)\;, & \bar{a} &=
  \frac{1}{\sqrt{2}}(x_1-\mathrm{i} x_2)\;, \nonumber
  \\*
  \frac{\partial}{\partial a} &= \frac{1}{\sqrt{2}}(\partial_1 -
  \mathrm{i} \partial_2 )\;, & \frac{\partial}{\partial \bar{a}} &=
  \frac{1}{\sqrt{2}}(\partial_1 + \mathrm{i} \partial_2 )\;.
\end{align}
For any $f \in \mathbb{R}^2_\theta$ we have
\begin{align}
  (a \star f)(x) &= a(x) f(x) + \frac{\theta_1}{2} \frac{\partial
    f}{\partial \bar{a}}(x)\;, & (f \star a)(x) &= a(x) f(x) -
  \frac{\theta_1}{2} \frac{\partial f}{\partial \bar{a}}(x)\;, \nonumber
  \\*
  (\bar{a} \star f)(x) &= \bar{a}(x) f(x) - \frac{\theta_1}{2}
  \frac{\partial f}{\partial a}(x)\;, & (f \star \bar{a})(x) &=
  \bar{a}(x) f(x) + \frac{\theta_1}{2} \frac{\partial f}{\partial
    a}(x)\;.
\end{align}
This implies $\bar{a}^{\star m} \star f_0=2^m \bar{a}^m f_0$, $f_0
\star a^{\star n} =2^n a^n f_0$ and
\begin{align}
  a \star \bar{a}^{\star m} \star f_0 &= \left\{\begin{array}{cl}
      m\theta_1 (\bar{a}^{\star (m-1)} \star f_0) & \text{ for } m\geq 1
      \\
      0 & \text{ for } m =0
\end{array}\right.
\nonumber
\\*
f_0 \star a^{\star n} \star \bar{a} &= \left\{\begin{array}{cl} n
    \theta_1 (f_0 \star a^{\star (n-1)})\;\, & \text{ for } n\geq 1
    \\
    0 & \text{ for } n =0
\end{array}\right.
\label{faa}
\end{align}
where $a^{\star n} = a \star a \star \dots \star a$ ($n$ factors) and
similarly for $\bar{a}^{\star m}$. Now, defining
\begin{align}
  f_{mn} &:=\frac{1}{\sqrt{n! m! \,\theta_1^{m+n}}} \, \bar{a}^{\star m}
  \star f_0 \star a^{\star n}
\label{fmn}
\\*
& = \frac{1}{\sqrt{n! m! \,\theta_1^{m+n}}} \sum_{k=0}^{\min(m,n)}
(-1)^k \binom{m}{k} \binom{n}{k} \,k! \,2^{m+n-2k}\, \theta_1^k
\,\bar{a}^{m-k} \,a^{n-k} f_0\;, \nonumber
\end{align}
(the second line is proved by induction) it follows from (\ref{faa})
and (\ref{ff}) that
\begin{align}
  (f_{mn} \star f_{kl})(x) = \delta_{nk} f_{ml}(x)\;.
\label{fprod}
\end{align}
The multiplication rule (\ref{fprod}) identifies the $\star$-product
with the ordinary matrix product:
\begin{align}
  a(x) &= \sum_{m,n=0}^\infty a_{mn} f_{mn}(x)\;,& b(x) &=
  \sum_{m,n=0}^\infty b_{mn} f_{mn}(x) \nonumber
  \\*
  \Rightarrow\quad (a\star b)(x) &= \sum_{m,n=0}^\infty (ab)_{mn}
  f_{mn}(x)\;, & (ab)_{mn} &= \sum_{k=0}^\infty a_{mk} b_{kn}\;.
\label{fprodmat}
\end{align}
In order to describe elements of $\mathbb{R}^2_\theta$ the sequences
$\{a_{mn}\}$ must be of rapid decay \cite{Gracia-Bondia:1987kw}:
\begin{align}
  \sum_{m,n=0}^\infty a_{mn} f_{mn} \in \mathbb{R}^2_\theta \qquad
  \text{iff} \quad \sum_{m,n=0}^\infty \big((2m{+}1)^{2k}(2n{+}1)^{2k}
  |a_{mn}|^2\big)^{\frac{1}{2}} < \infty \quad \text{for all } k\;.
\end{align}
Finally, using (\ref{ff}), the trace property of the integral and
(\ref{faa}) we compute
\begin{align}
  \int d^2 x \, f_{mn}(x) &= \frac{1}{\sqrt{m! n!\, \theta_1^{m+n}}}
  \int d^2x\, \big( \bar{a}^{\star m} \star f_0 \star f_0 \star
  a^{\star n} \big)(x) 
= \delta_{mn} \int d^2x f_0(x) 
\nonumber
\\*
&= 2 \pi \theta_1 \delta_{mn}\;.
\label{intfmn}
\end{align}

The functions $f_{mn}$ with $m,n <\mathcal{N}$ provide a cut-off both
in position and momentum space. Passing to radial coordinates
$x_1=\rho \cos \varphi$, $x_2=\rho \sin \varphi$ we can compare
(\ref{fmn}) with the expansion of Laguerre polynomials
\cite[\S 8.970.1]{GR}:
\begin{align}
  f_{mn}(\rho,\varphi) &= 2(-1)^m \sqrt{\tfrac{m!}{n!}}
  \mathrm{e}^{\mathrm{i}\varphi(n-m)} \Big(\sqrt{\tfrac{2}{\theta_1}}
  \rho\Big)^{n-m} L^{n-m}_m(\tfrac{2}{\theta_1}\rho^2)
  \,\mathrm{e}^{-\frac{\rho^2}{\theta_1}} \;.
\label{frho}
\end{align}
The function $L^\alpha_m(z) z^{\alpha/2} \mathrm{e}^{-z/2}$ is rapidly
decreasing beyond the last maximum $(z^\alpha_m)_{\text{max}}$. One
finds numerically $(z^\alpha_m)_{\text{max}} < 2\alpha + 4 m$ and thus
the radial cut-off
\begin{align}
\rho_{\text{max}} \approx \sqrt{2 \theta_1 \mathcal{N}} 
\qquad  \text{for $m,n < \mathcal{N}$}\;.
\end{align}

On the other hand, for $p_1=-p\sin\psi$, $p_2=p\cos\psi$ we compute
with (\ref{frho}), \cite[\S 8.411.1]{GR} and \cite[\S 7.421.5]{GR}
\begin{align}
\tilde{f}(p,\psi) 
&:= \int_0^\infty \!\!\!\rho\, d\rho \int_{0}^{2\pi} \!\!d\varphi \;
\mathrm{e}^{\mathrm{i} p\rho \sin(\varphi-\psi)}  f_{mn}(\rho,\varphi) 
\nonumber
\\*
&= 4\pi(-1)^n \sqrt{\tfrac{m!}{n!}}\mathrm{e}^{\mathrm{i}\psi(n-m)}
\int_0^\infty \!\!\!\rho d\rho \,
   \Big(\sqrt{\tfrac{2}{\theta_1}}
  \rho\Big)^{n-m} L^{n-m}_m(\tfrac{2}{\theta_1}\rho^2)
  J_{n-m}(\rho p)\,\mathrm{e}^{-\frac{\rho^2}{\theta_1}} 
\nonumber
\\*
&= 2\pi\theta_1 \sqrt{\tfrac{m!}{n!}}\mathrm{e}^{\mathrm{i}(\psi+\pi)(n-m)} 
   \Big(\sqrt{\tfrac{\theta_1}{2}}\,p \Big)^{n-m} 
L^{n-m}_m(\tfrac{\theta_1}{2}p^2)
\mathrm{e}^{-\frac{\theta_1}{4}p^2}\;. 
\end{align}
We thus have 
\begin{align}
  p_{\text{max}} \approx \sqrt{\frac{8 \mathcal{N}}{\theta_1}} 
\qquad  \text{for $m,n < \mathcal{N}$}\;.
\label{pmax}
\end{align}

\section{Asymptotic behaviour of the propagator}
\label{appB}

\begin{figure}[hb!]
\begin{picture}(160,135)
\put(-20,-32){\epsfig{file=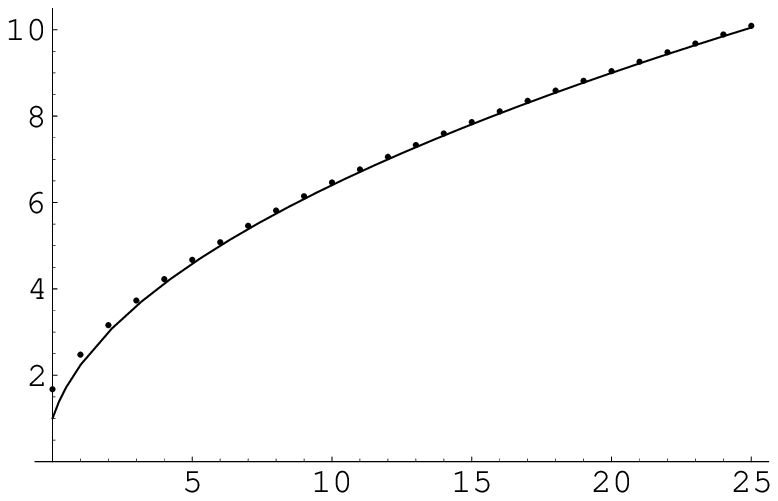,scale=0.6}}
\put(35,-32){\epsfig{file=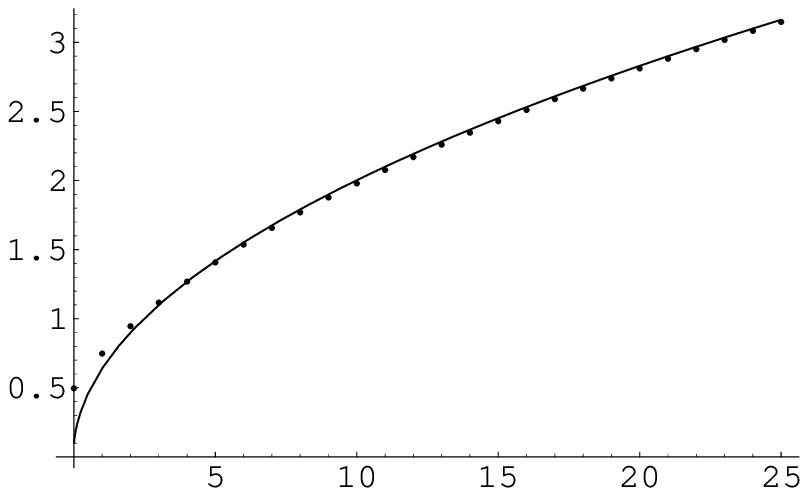,scale=0.6}}
\put(90,-32){\epsfig{file=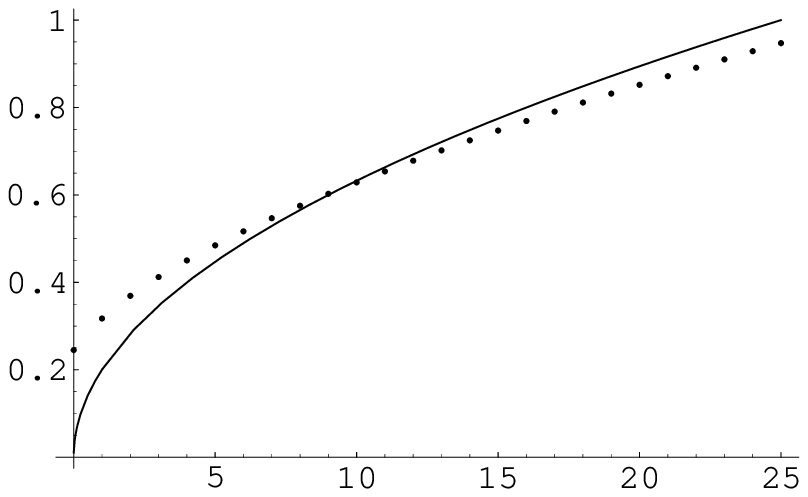,scale=0.6}}
\put(22,106){\mbox{\scriptsize$\omega=1\,,~\mu_0^2=1$}}
\put(76,106){\mbox{\scriptsize$\omega=1\,,~\mu_0^2=0.1$}}
\put(130,106){\mbox{\scriptsize$\omega=1\,,~\mu_0^2=0.01$}}
\put(-20,-66){\epsfig{file=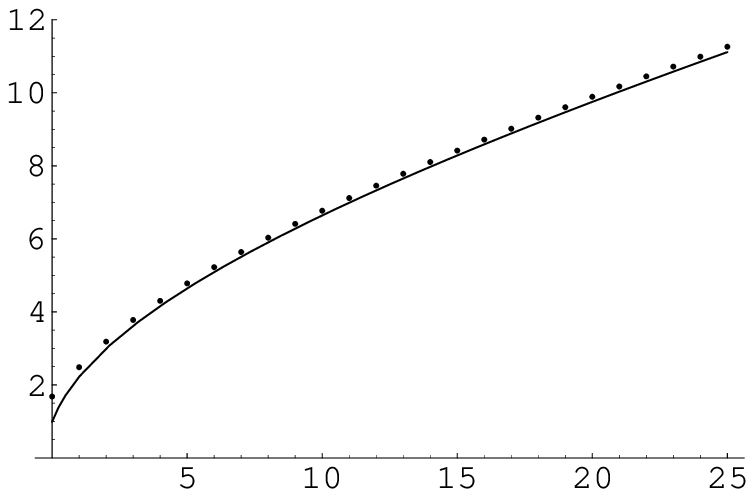,scale=0.6}}
\put(35,-66){\epsfig{file=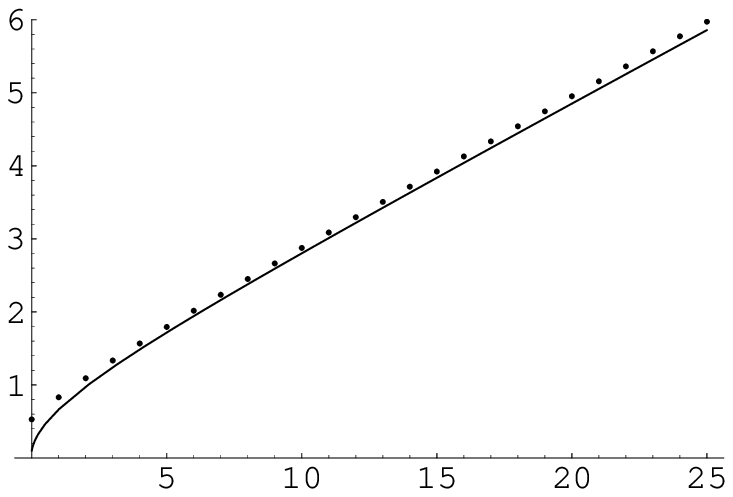,scale=0.6}}
\put(90,-66){\epsfig{file=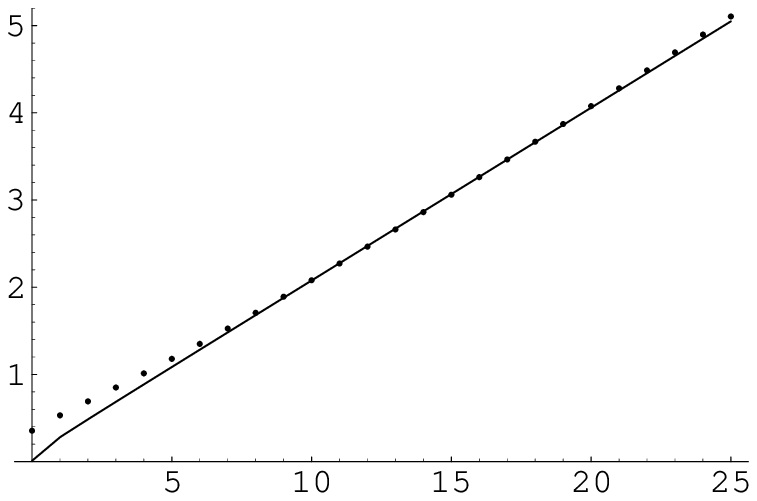,scale=0.6}}
\put(22,72){\mbox{\scriptsize$\omega=0.99\,,~\mu_0^2=1$}}
\put(76,72){\mbox{\scriptsize$\omega=0.99\,,~\mu_0^2=0.1$}}
\put(130,72){\mbox{\scriptsize$\omega=0.99\,,~\mu_0^2=0.01$}}
\put(-20,-100){\epsfig{file=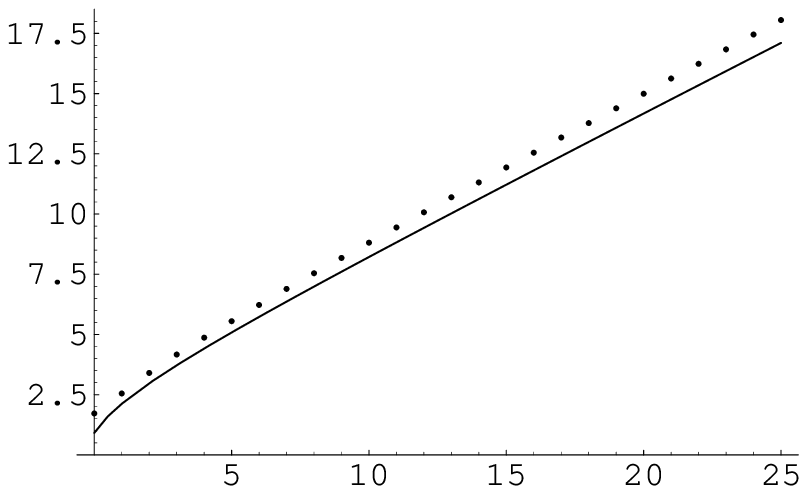,scale=0.6}}
\put(35,-100){\epsfig{file=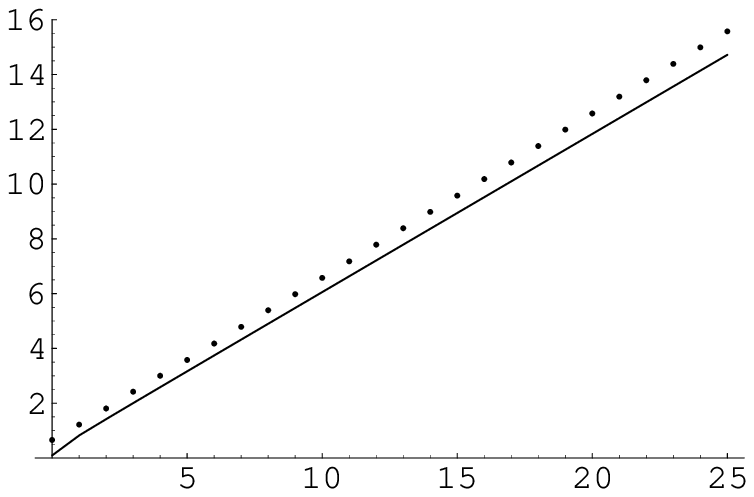,scale=0.6}}
\put(90,-100){\epsfig{file=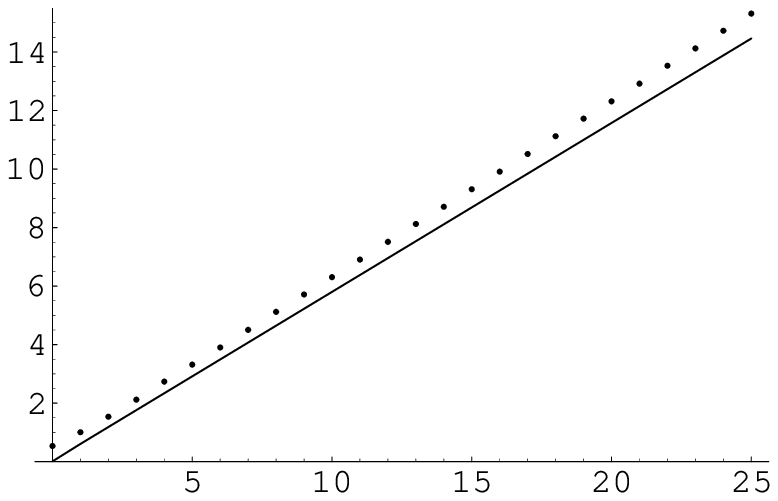,scale=0.6}}
\put(22,38){\mbox{\scriptsize$\omega=0.9\,,~\mu_0^2=1$}}
\put(76,38){\mbox{\scriptsize$\omega=0.9\,,~\mu_0^2=0.1$}}
\put(130,38){\mbox{\scriptsize$\omega=0.9\,,~\mu_0^2=0.01$}}
\put(-20,-134){\epsfig{file=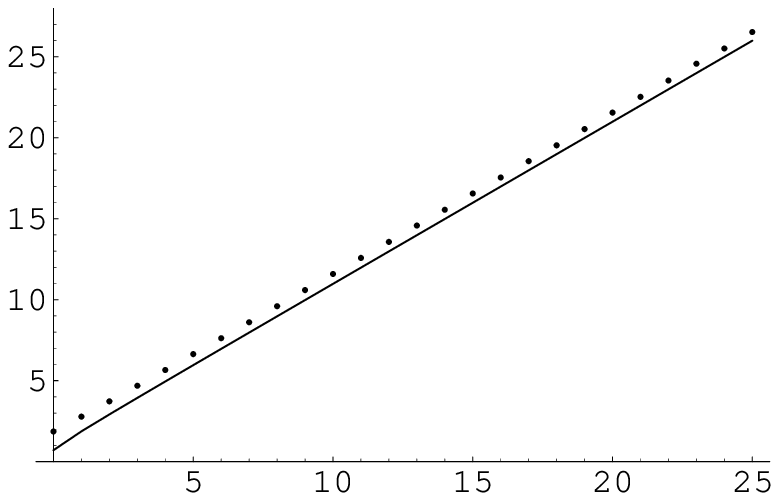,scale=0.6}}
\put(35,-134){\epsfig{file=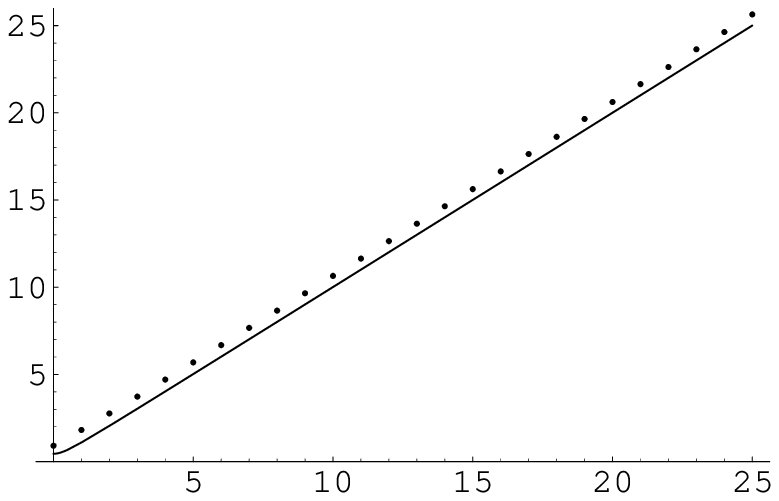,scale=0.6}}
\put(90,-134){\epsfig{file=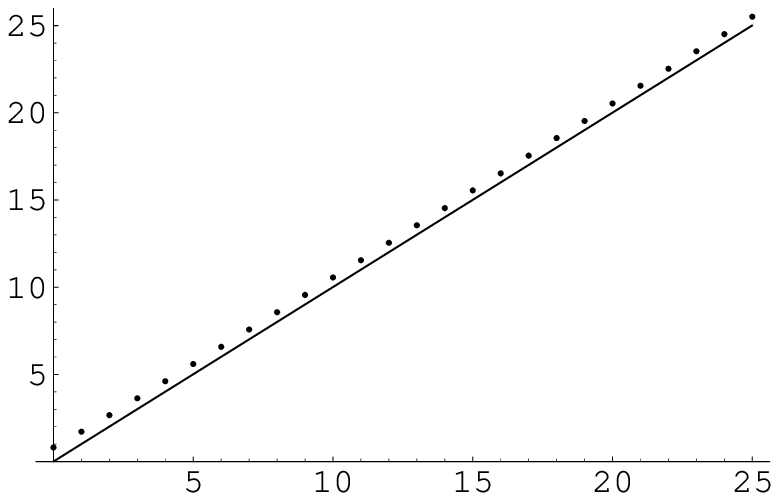,scale=0.6}}
\put(22,4){\mbox{\scriptsize$\omega=0.5\,,~\mu_0^2=1$}}
\put(76,4){\mbox{\scriptsize$\omega=0.5\,,~\mu_0^2=0.1$}}
\put(130,4){\mbox{\scriptsize$\omega=0.5\,,~\mu_0^2=0.01$}}
\end{picture}
\caption{$(\max \Delta_{nm;kl}^{\mathcal{C}}(\mu_0^2,\mu^2))^{-1}$
  compared with
  $((\mu_0^4+2\mu_0^2\mu^2\mathcal{C}+4\mu^4(1{-}\omega)\mathcal{C}^2
  )/(3-2\omega))^{\frac{1}{2}}$, both plotted over $\mathcal{C}$, for various
  parameters $\omega$ and $\mu_0^2$. We have normalised $\mu^2=1$.}
\label{fig-Form1}
\end{figure}

\begin{figure}[h!t]
\begin{picture}(160,35)
\put(-20,-134){\epsfig{file=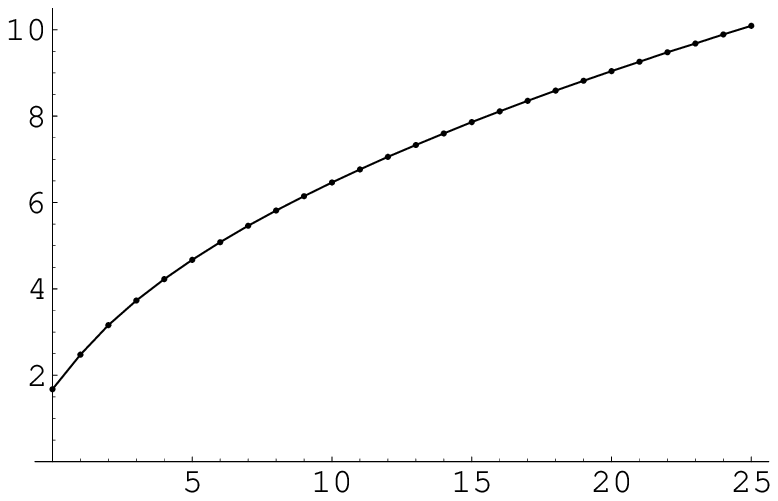,scale=0.6}}
\put(35,-134){\epsfig{file=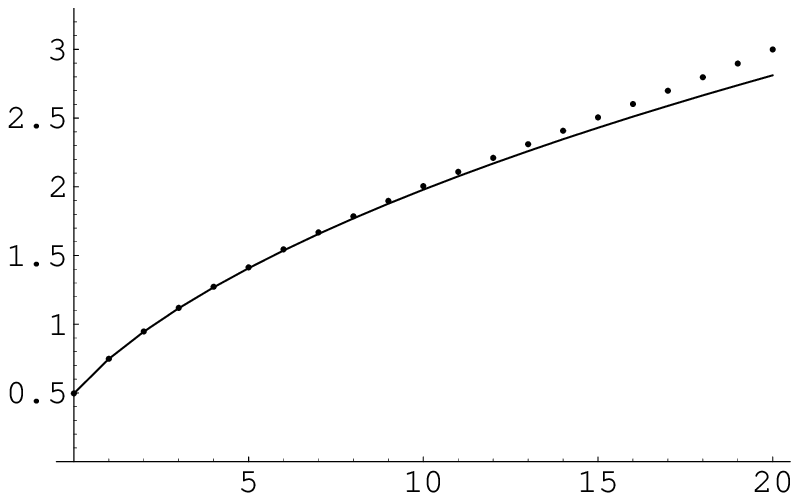,scale=0.6}}
\put(90,-134){\epsfig{file=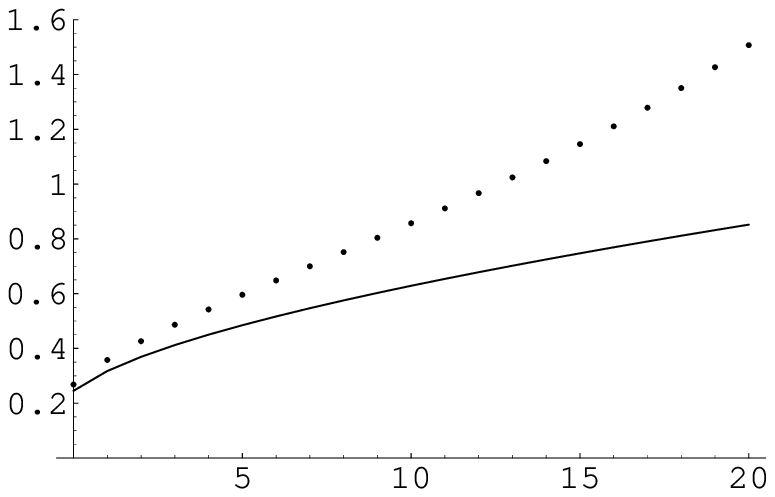,scale=0.6}}
\put(22,4){\mbox{\scriptsize$\omega=1\,,~\mu_0^2=1$}}
\put(76,4){\mbox{\scriptsize$\omega=1\,,~\mu_0^2=0.1$}}
\put(130,4){\mbox{\scriptsize$\omega=1\,,~\mu_0^2=0.01$}}
\end{picture}
\caption{$(\max \Delta_{nm;kl}^{\mathcal{C}}(\mu_0^2,\mu^2))^{-1}$
  for $\omega=1$ computed with (\ref{DeltaN}) and $\mathcal{N}=55$
  (dots) and with (\ref{Deltaexact}) (solid curve), both plotted
  over $\mathcal{C}$, for various parameters $\mu_0^2$. We have
  normalised $\mu^2=1$.  It is apparent that (\ref{DeltaN}) converges
  badly for large $\frac{\mu^2}{\mu_0^2}$.}
\label{fig-Form1-lag}
\end{figure}

The crucial question for renormalisation is how fast the propagator
$\Delta^K_{nm;lk}(\mu^2,\mu_0^2)$ and a certain summation over its
indices decay if the indices $m,n,k,l$ become large. We need two
asymptotic formulae which we deduce from the numerical
evaluation of the propagator for a representative class of parameters.
These formulae involve the cut-off propagator
\begin{align}
  \Delta_{nm;lk}^{\mathcal{C}} := \left\{ \begin{array}{ll}
      \Delta_{nm;lk} \qquad & \text{for } \mathcal{C} \leq
      \max(m,n,k,l) \leq 2\mathcal{C}\;,
      \\
      0 & \text{otherwise\;,}
\end{array}\right.
\end{align}
which is the restriction of $\Delta_{nm;lk}$ to the support of the
cut-off propagator $\Delta_{nm;lk}^K(\Lambda)$ appearing in the
Polchinski equation, with $\mathcal{C}=\frac{\Lambda^2}{\mu^2}$.
\\[\bigskipamount]
{\bf Formula 1:}
\begin{align}
  \max_{m,n,k,l}\Big( \Delta_{nm;lk}^{\mathcal{C}}(\mu^2,\mu_0^2)
  \Big) \approx \sqrt{\frac{3-2\omega}{\mu^4_0 
+ 4 \mu_0^2\mu^2 \mathcal{C} + 4 \mu^4 (1{-}\omega) 
\mathcal{C}^2}}\,\delta_{m+k,n+l}\;.
\label{Form1}
\end{align}
We demonstrate in Figure~\ref{fig-Form1}% 
that $(\max
\Delta_{nm;lk}^{\mathcal{C}})^{-1}$ is asymptotically reproduced by
$((\mu^4_0 + 4 \mu_0^2\mu^2 \mathcal{C} + 4 \mu^4 (1{-}\omega)
\mathcal{C}^2)/(3-2\omega))^{\frac{1}{2}}$.
We have evaluated the formula (\ref{DeltaN}) for the propagator with
$\mathcal{N}=55$. An exception is $\Delta_{nm;lk}$ for $\omega=1$ and
$\mu^2\gg \mu_0^2$. Here the choice $\mathcal{N}=55$ in (\ref{DeltaN})
is too small, and we have used the numerical evaluation of
(\ref{Deltaexact}) instead. We compare the outcome of (\ref{DeltaN})
for $\omega=1$ and (\ref{Deltaexact}) for various values of $\mu_0^2$
in Figure~\ref{fig-Form1-lag}.
\begin{figure}[h!b]
\begin{picture}(160,140)
\put(-20,-32){\epsfig{file=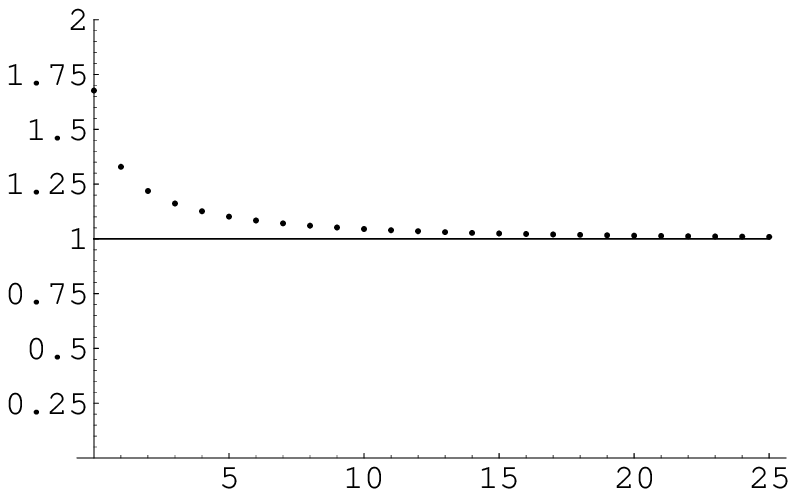,scale=0.6}}
\put(35,-32){\epsfig{file=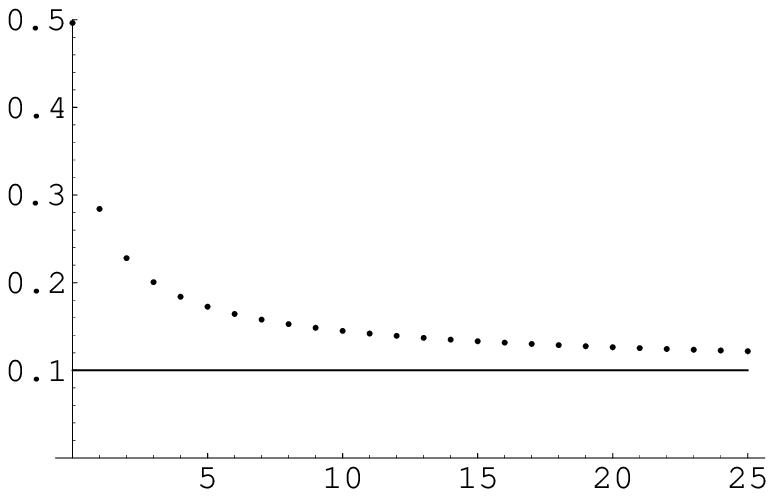,scale=0.6}}
\put(90,-32){\epsfig{file=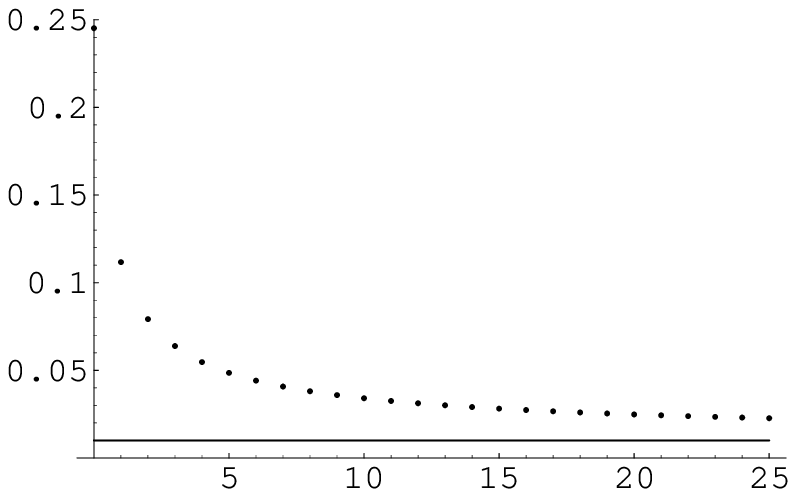,scale=0.6}}
\put(22,106){\mbox{\scriptsize$\omega=1\,,~\mu_0^2=1$}}
\put(76,126){\mbox{\scriptsize$\omega=1\,,~\mu_0^2=0.1$}}
\put(130,126){\mbox{\scriptsize$\omega=1\,,~\mu_0^2=0.01$}}
\put(-20,-66){\epsfig{file=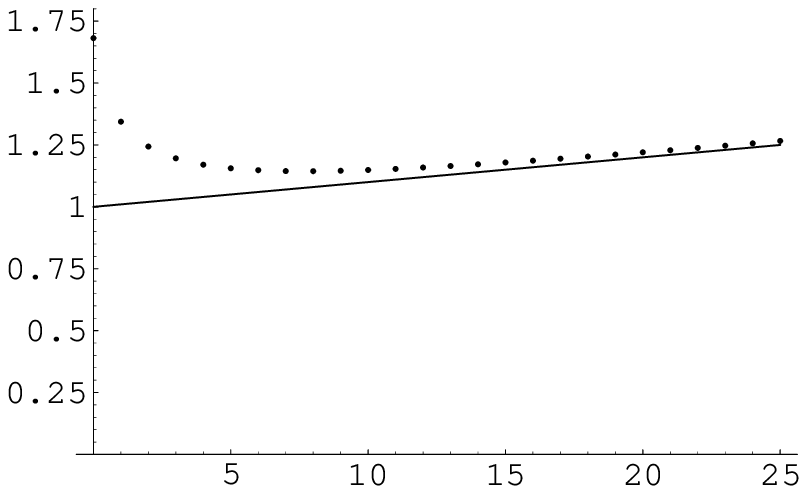,scale=0.6}}
\put(35,-66){\epsfig{file=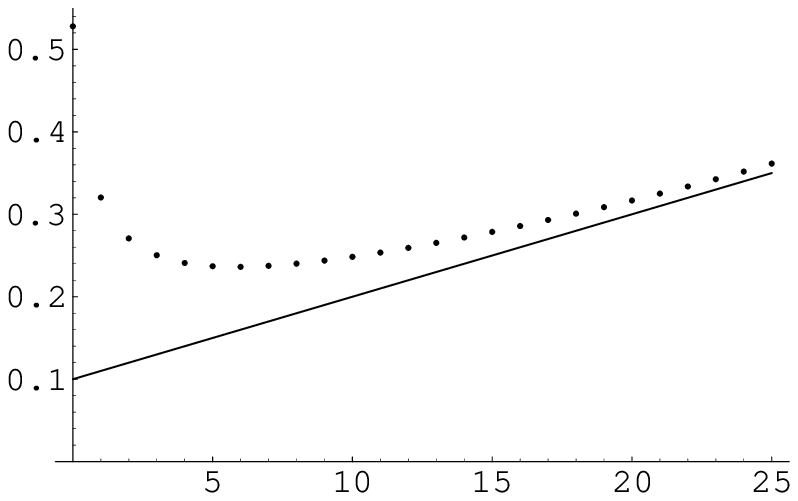,scale=0.6}}
\put(90,-66){\epsfig{file=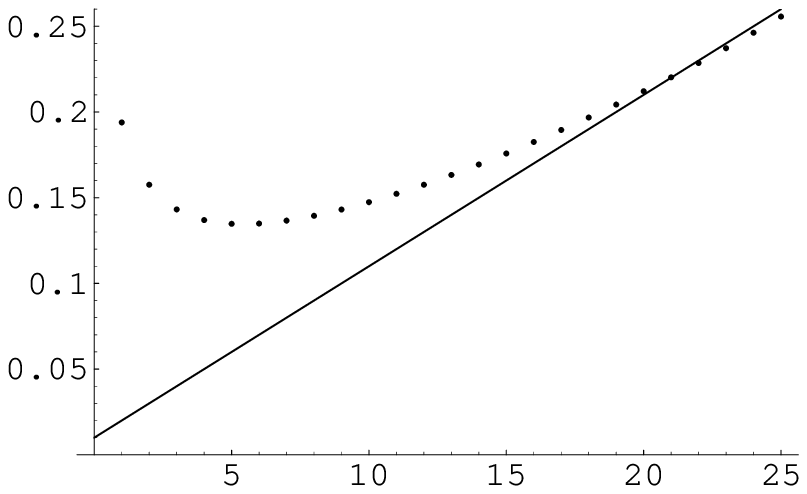,scale=0.6}}
\put(22,72){\mbox{\scriptsize$\omega=0.99\,,~\mu_0^2=1$}}
\put(76,72){\mbox{\scriptsize$\omega=0.99\,,~\mu_0^2=0.1$}}
\put(130,72){\mbox{\scriptsize$\omega=0.99\,,~\mu_0^2=0.01$}}
\put(-20,-100){\epsfig{file=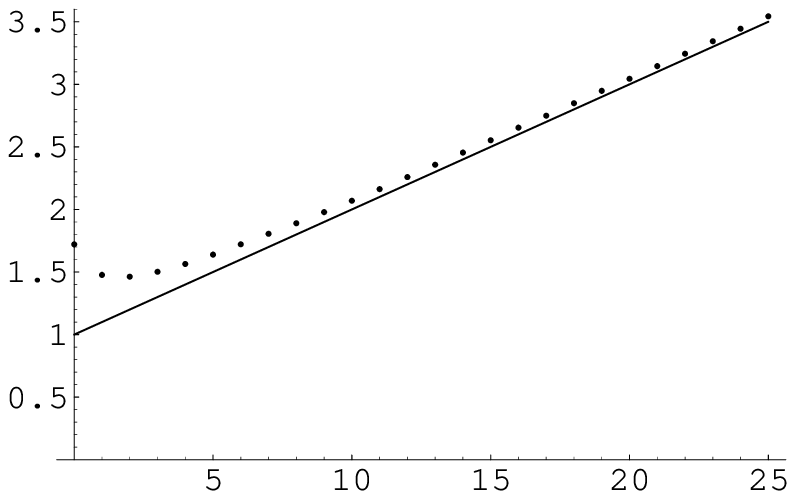,scale=0.6}}
\put(35,-100){\epsfig{file=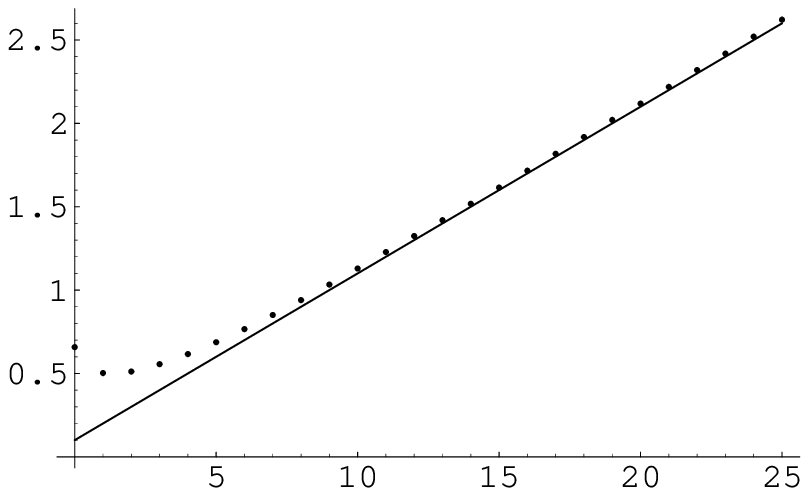,scale=0.6}}
\put(90,-100){\epsfig{file=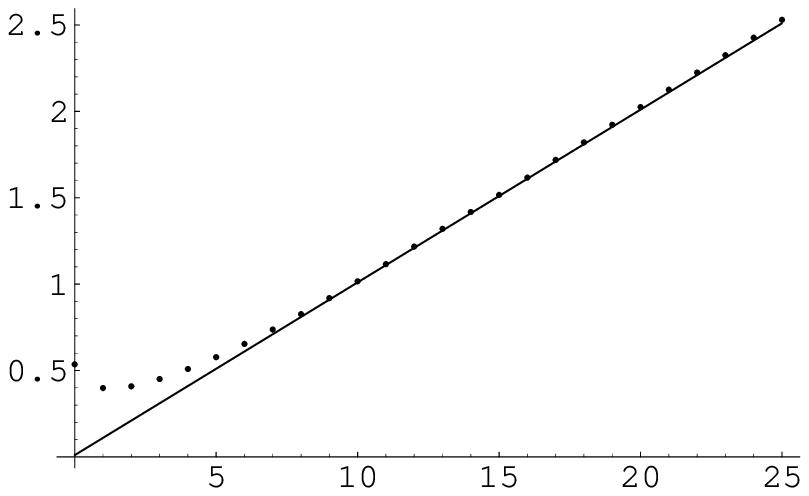,scale=0.6}}
\put(22,38){\mbox{\scriptsize$\omega=0.9\,,~\mu_0^2=1$}}
\put(76,38){\mbox{\scriptsize$\omega=0.9\,,~\mu_0^2=0.1$}}
\put(130,38){\mbox{\scriptsize$\omega=0.9\,,~\mu_0^2=0.01$}}
\put(-20,-134){\epsfig{file=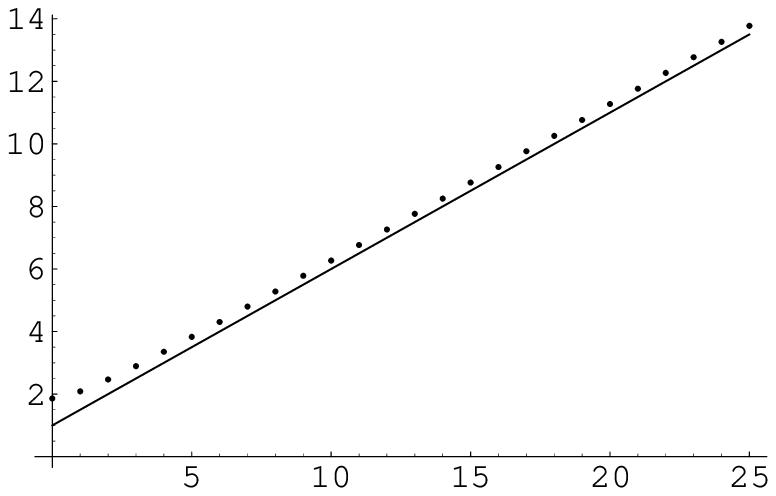,scale=0.6}}
\put(35,-134){\epsfig{file=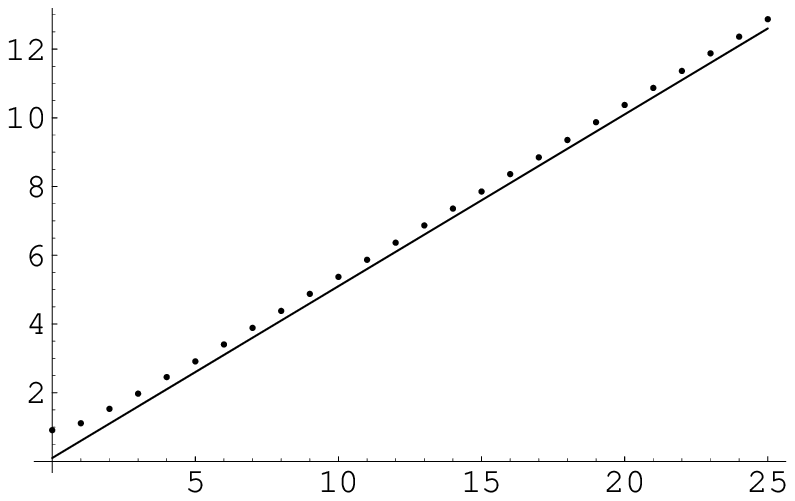,scale=0.6}}
\put(90,-134){\epsfig{file=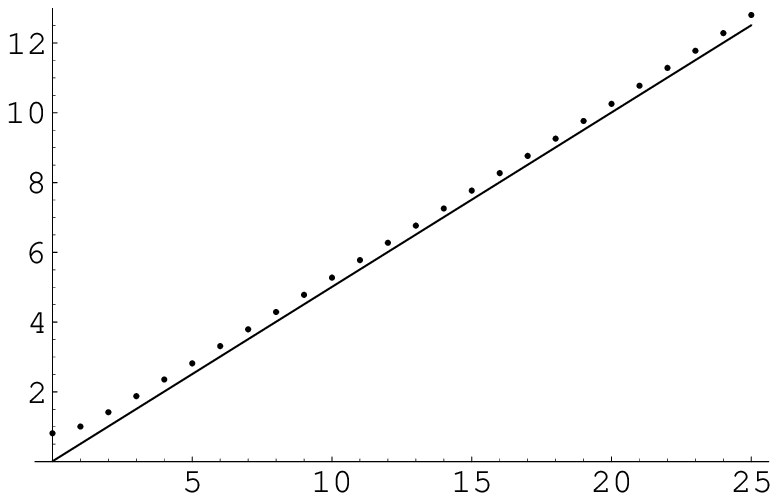,scale=0.6}}
\put(22,4){\mbox{\scriptsize$\omega=0.5\,,~\mu_0^2=1$}}
\put(76,4){\mbox{\scriptsize$\omega=0.5\,,~\mu_0^2=0.1$}}
\put(130,4){\mbox{\scriptsize$\omega=0.5\,,~\mu_0^2=0.01$}}
\end{picture}
\caption{$( \max_{n} \sum_{k} \max_{m,l}
  \Delta_{nm;lk}^{\mathcal{C}})^{-1}$ compared with
  $\mu_0^2+\mu^2(1{-}\omega)\mathcal{C}$, both plotted over
  $\mathcal{C}$, for various parameters $\omega$ and $\mu_0^2$. We have
  normalised $\mu^2=1$.}
\label{fig-Form2}
\end{figure}

\clearpage
\noindent
{\bf Formula 2:}
\begin{align}
  \max_{n} \sum_{k} \max_{m,l}
  \Delta_{nm;lk}^{\mathcal{C}}(\mu^2,\mu_0^2) \approx
  \frac{1}{\mu_0^2 + \mu^2 (1{-}\omega) \mathcal{C}}\;.
\label{Form2}
\end{align}
We demonstrate in Figure~\ref{fig-Form2} that $( \max_{n} \sum_{k}
\max_{m,l} \Delta_{nm;lk}^{\mathcal{C}})^{-1}$ is asymptotically given
by $\mu^2_0 + \mu^2 (1{-}\omega) \mathcal{C}$. We have evaluated the
formula (\ref{DeltaN}) for the propagator with $\mathcal{N}=55$,
except for $\omega=1$ and $\mu^2\gg \mu_0^2$, where (\ref{Deltaexact})
is used. The crucial observation is that for $\omega=1$ the function
$\max_{n} \sum_{k} \max_{m,l} \Delta_{nm;lk}^{\mathcal{C}}$ is
\emph{increasing} with $\mathcal{C}$ so that $\lim_{\mathcal{C}\to
  \infty} \max_{n} \sum_{k} \max_{m,l}\Delta_{nm;lk}^{\mathcal{C}} =
\mu_0^{-2}>0$.

\begin{figure}[h!]
\begin{picture}(160,140)
\put(-20,-32){\epsfig{file=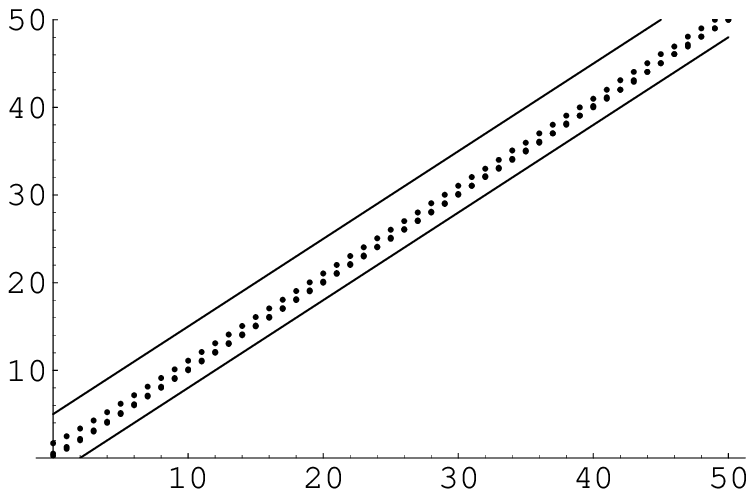,scale=0.6}}
\put(35,-32){\epsfig{file=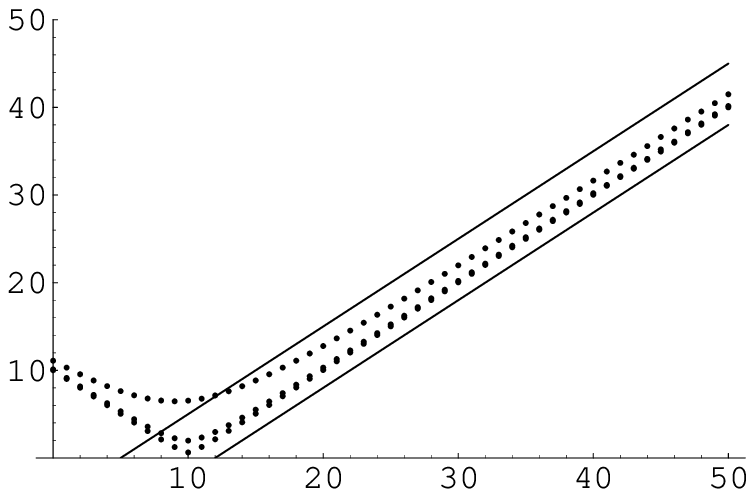,scale=0.6}}
\put(90,-32){\epsfig{file=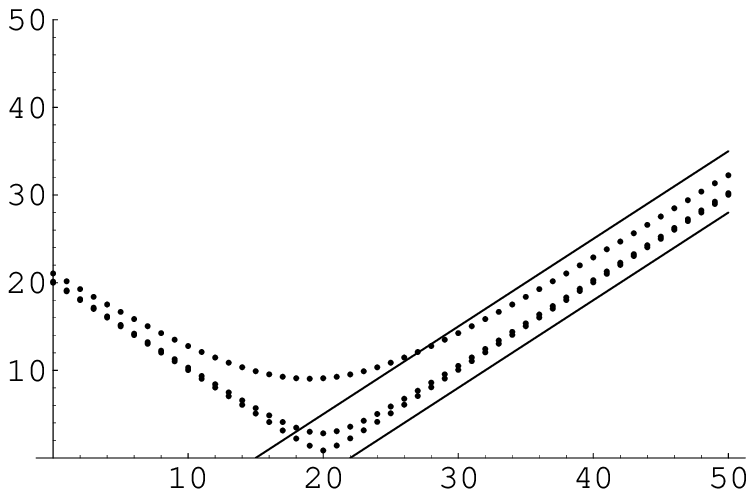,scale=0.6}}
\put(22,106){\mbox{\scriptsize$\omega=1\,,~m=0$}}
\put(66,123){\mbox{\scriptsize$\omega=1\,,~m=10$}}
\put(120,123){\mbox{\scriptsize$\omega=1\,,~m=20$}}
\put(-20,-66){\epsfig{file=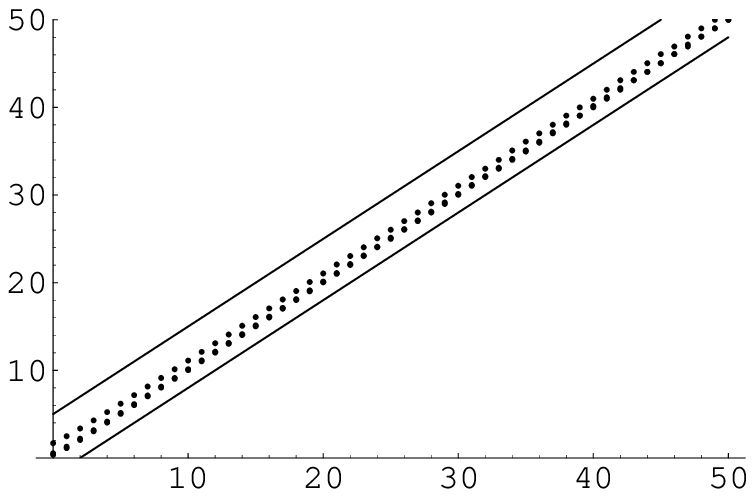,scale=0.6}}
\put(35,-66){\epsfig{file=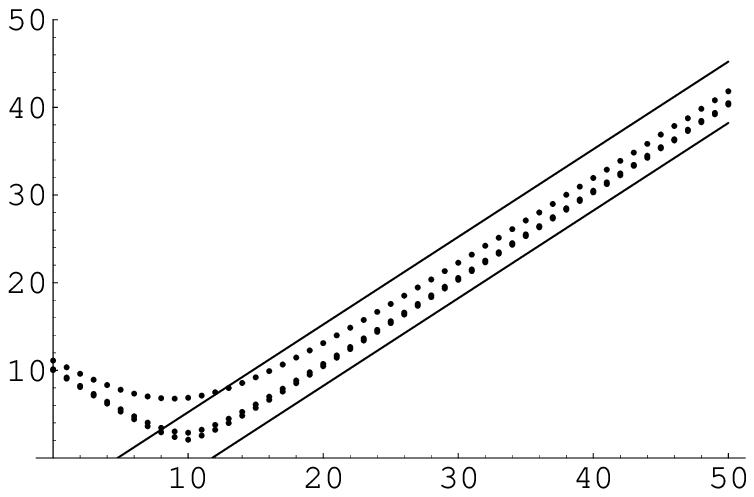,scale=0.6}}
\put(90,-66){\epsfig{file=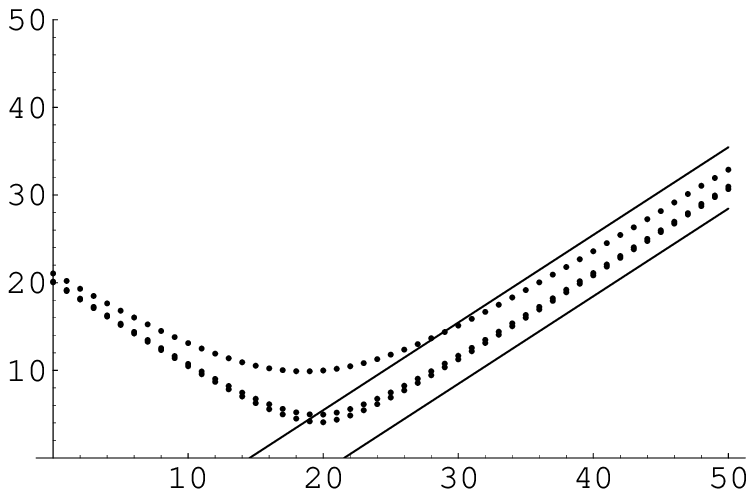,scale=0.6}}
\put(22,72){\mbox{\scriptsize$\omega=0.99\,,~m=0$}}
\put(66,89){\mbox{\scriptsize$\omega=0.99\,,~m=10$}}
\put(120,89){\mbox{\scriptsize$\omega=0.99\,,~m=20$}}
\put(-20,-100){\epsfig{file=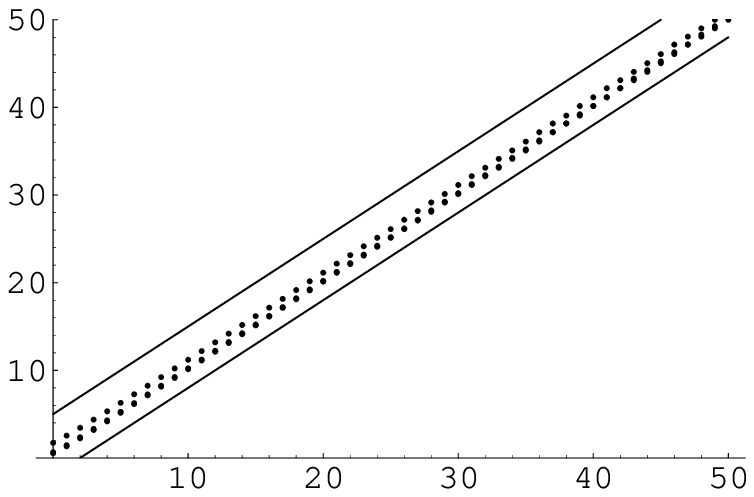,scale=0.6}}
\put(35,-100){\epsfig{file=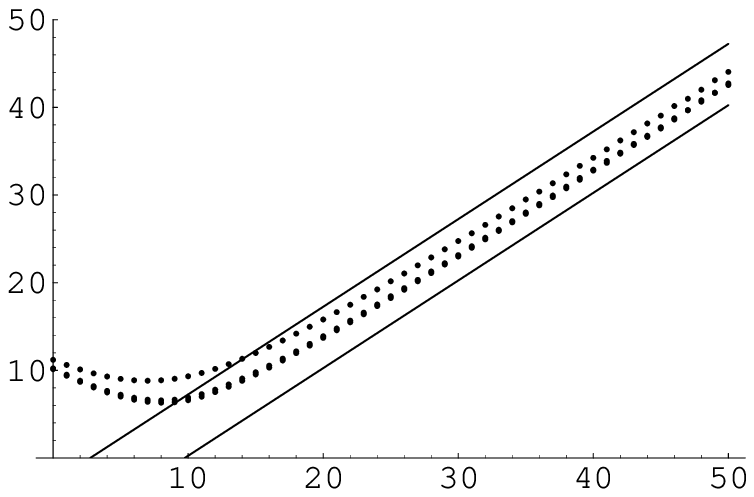,scale=0.6}}
\put(90,-100){\epsfig{file=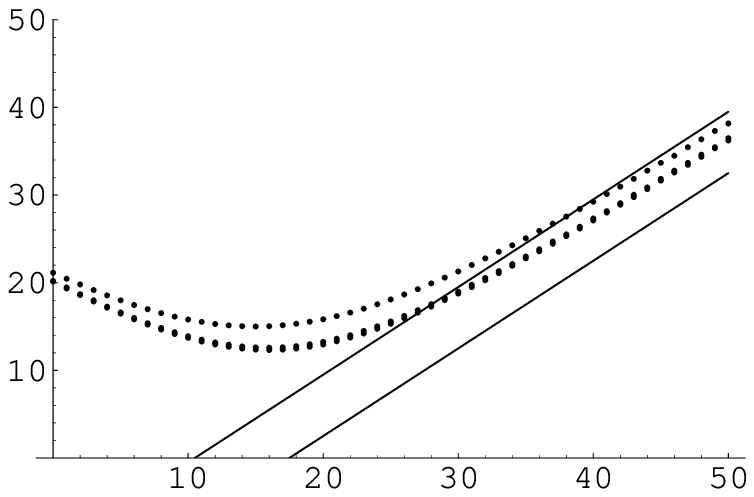,scale=0.6}}
\put(22,38){\mbox{\scriptsize$\omega=0.9\,,~m=0$}}
\put(66,55){\mbox{\scriptsize$\omega=0.9\,,~m=10$}}
\put(120,55){\mbox{\scriptsize$\omega=0.9\,,~m=20$}}
\put(-20,-134){\epsfig{file=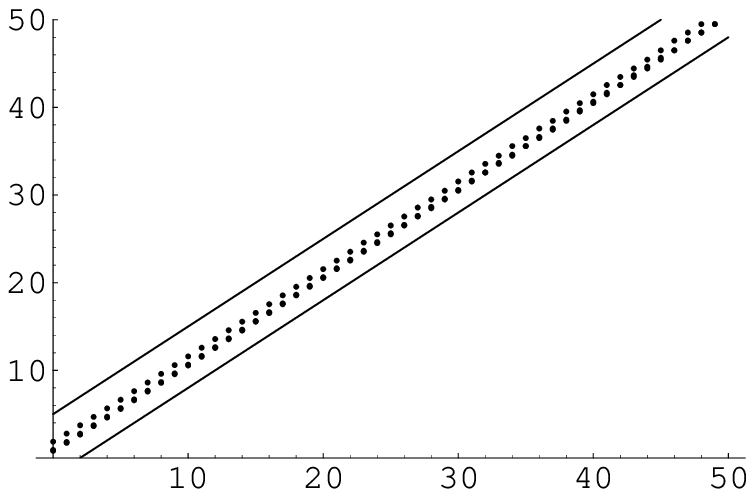,scale=0.6}}
\put(35,-134){\epsfig{file=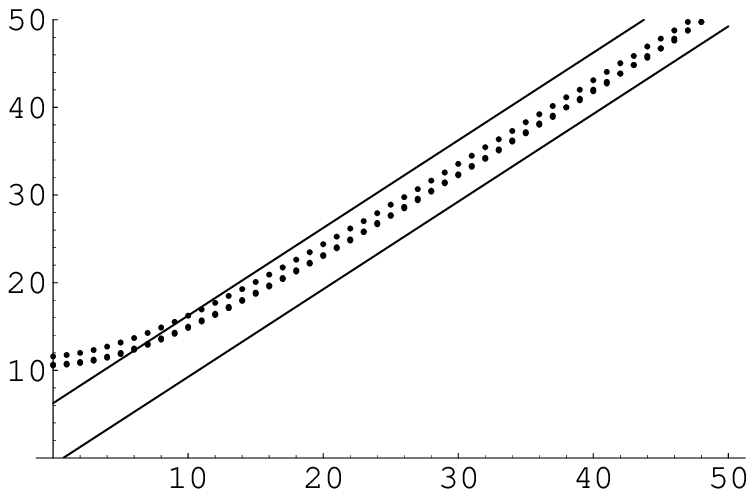,scale=0.6}}
\put(90,-134){\epsfig{file=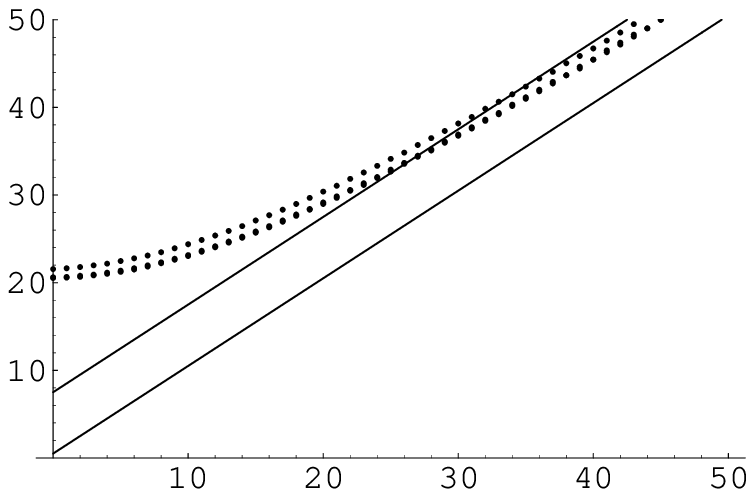,scale=0.6}}
\put(22,4){\mbox{\scriptsize$\omega=0.5\,,~m=0$}}
\put(76,4){\mbox{\scriptsize$\omega=0.5\,,~m=10$}}
\put(130,4){\mbox{\scriptsize$\omega=0.5\,,~m=20$}}
\end{picture}
\caption{$(\mu^2 \Delta_{nm;mn})^{-1}$ compared with 
$n-\frac{9\omega-5}{4}m+5$ and $n-\frac{9\omega-5}{4}m-2$, both
plotted over $n$, for various parameters $\omega$ and $m$. The dots
show $(\mu^2 \Delta_{nm;mn})^{-1}$ for three values $\mu_0 =\mu$ (upper dots), 
$\mu_0 =0.1\mu$ and $\mu_0 =0.01 \mu$ (lower dots) of the mass.}
\label{fig3}
\end{figure}
Finally, for the verification of (\ref{consistency}) we need 
\begin{align}
\frac{1}{\mu^2(n-\frac{9\omega-5}{4}m+5)} <
\Delta_{nm;mn}(\mu^2,\mu_0^2) < \frac{1}{\mu^2(n-\frac{9\omega-5}{4}m+5)-2}
\qquad \text{for } m \ll n \;,
\label{asDelta}  
\end{align}
independent of $\mu_0$. We compare in Figure~\ref{fig3} the inverse of
the matrix element $\mu^2 \Delta_{nm;mn}(\mu^2,\mu_0^2)$ of the propagator
with the asymptotics
$n-\frac{9\omega-5}{4}m\big\{\genfrac{}{}{0pt}{1}{+5}{-2}$.

\end{appendix}

\begin{thebibliography}{99}

%\cite{Minwalla:1999px}
\bibitem{Minwalla:1999px}
S.~Minwalla, M.~Van Raamsdonk and N.~Seiberg,
``Noncommutative perturbative dynamics,''
JHEP {\bf 0002} (2000) 020
[arXiv:hep-th/9912072].
%%CITATION = HEP-TH 9912072;%%

%\cite{Chepelev:1999tt}
\bibitem{Chepelev:1999tt} I.~Chepelev and R.~Roiban, ``Renormalization
  of quantum field theories on noncommutative $\mathbb{R}^d$.  I:
  Scalars,'' JHEP {\bf 0005} (2000) 037 [arXiv:hep-th/9911098].
%%CITATION = HEP-TH 9911098;%%

%\cite{Chepelev:2000hm}
\bibitem{Chepelev:2000hm} I.~Chepelev and R.~Roiban, ``Convergence
  theorem for non-commutative Feynman graphs and renormalization,''
  JHEP {\bf 0103} (2001) 001 [arXiv:hep-th/0008090].
%%CITATION = HEP-TH 0008090;%%

%\cite{Grosse:2003aj}
\bibitem{Grosse:2003aj}
H.~Grosse and R.~Wulkenhaar,
``Power-counting theorem for non-local matrix models and renormalisation,''
arXiv:hep-th/0305066.
%%CITATION = HEP-TH 0305066;%%

%\cite{Wilson:1973jj}
\bibitem{Wilson:1973jj}
K.~G.~Wilson and J.~B.~Kogut,
``The Renormalization Group And The Epsilon Expansion,''
Phys.\ Rept.\  {\bf 12} (1974) 75.
%%CITATION = PRPLC,12,75;%%

%\cite{Polchinski:1983gv}
\bibitem{Polchinski:1983gv}
J.~Polchinski,
``Renormalization And Effective Lagrangians,''
Nucl.\ Phys.\ B {\bf 231} (1984) 269.
%%CITATION = NUPHA,B231,269;%%

%\cite{Gracia-Bondia:1987kw}
\bibitem{Gracia-Bondia:1987kw} J.~M.~Gracia-Bond\'{\i}a and
  J.~C.~V\'arilly, ``Algebras Of Distributions Suitable For Phase
  Space Quantum Mechanics. 1,'' J.\ Math.\ Phys.\ {\bf 29} (1988) 869.
%%CITATION = JMAPA,29,869;%%

%\cite{gw3}
\bibitem{gw3} H.~Grosse and R.~Wulkenhaar, ``Renormalisation of
  $\phi^4$-theory on noncommutative $\mathbb{R}^4$ in the
  matrix base,'' in preparation.

\bibitem{GR}
I.~S.~Gradshteyn and I.~M.~Ryzhik, ``Tables of Series, Produces, and
Integrals. Sixth Edition,'' Academic Press, San Diego (2000).

%\cite{Becchi:2003dg}
\bibitem{Becchi:2003dg} C.~Becchi, S.~Giusto and C.~Imbimbo, ``The
  renormalization of non-commutative field theories in the limit of
  large non-commutativity,'' arXiv:hep-th/0304159.
%%CITATION = HEP-TH 0304159;%%

%\cite{Zimmermann:1984sx}
\bibitem{Zimmermann:1984sx}
W.~Zimmermann,
``Reduction In The Number Of Coupling Parameters,''
Commun.\ Math.\ Phys.\  {\bf 97} (1985) 211.
%%CITATION = CMPHA,97,211;%%

\end{thebibliography}
\end{document}